\documentstyle[aps]{revtex}

\begin{document}
\bibliographystyle{prsty}

%
%

\title{
Monte Carlo Methods for Equilibrium and Nonequilibrium Problems 
in Interfacial Electrochemistry
}

\author{
Gregory Brown$^{\dagger\ddagger}$, 
Per Arne Rikvold$^{\dagger\ddagger}$, 
S. J. Mitchell$^{\dagger\ddagger}$, and 
M. A.\ Novotny$^{\ddagger}$
}

\address{
$^\dagger$ Center for Materials Research and Technology and Department of Physics\\ 
Florida State University, Tallahassee, FL 32306-4350 \\
$^\ddagger$ Supercomputer Computations Research Institute\\
Florida State University, Tallahassee, FL 32306-4130
}

\maketitle

\begin{abstract}
We present a tutorial discussion of Monte Carlo methods for equilibrium and 
nonequilibrium problems in interfacial electrochemistry. The discussion is 
illustrated with results from simulations of three specific systems: 
bromine adsorption on silver (100), underpotential deposition of copper on 
gold (111), and electrodeposition of urea on platinum (100). 
\end{abstract}

%
%

\section{Introduction}
\label{sec:intro}

In the context of interfacial electrochemistry the bulk electrolyte
can conveniently be thought of as a large, homogeneous reservoir of
charge, adsorbate particles, and energy. Then one can focus on the
metal--electrolyte interface, which can be viewed as an essentially
two-dimensional system of ions and molecules at or near a solid
surface. Consequently, interfacial electrochemistry can profitably be
thought of as a branch of surface science. It is therefore natural
that a number of experimental and theoretical methods originally
associated with more traditional branches of surface science now are
routinely applied in interfacial electrochemistry, including several
that are discussed in this volume.

On the theoretical side, one such method is computational lattice-gas
modeling.  This method is particularly well suited to the study of
fluctuations and phase transitions in systems characterized by
specific chemisorption at well defined sites on the surface. Different
aspects of this approach have been reviewed several times by others
\cite{BLUM90,EINS91,EINS95,BLUM96}, as well as by one of the present
authors \cite{RIKV91B,RIKV95,RIKV96}. Here, it will only be described
briefly. Monte Carlo simulation of lattice-gas models as a tool for
understanding interfacial electrochemistry is the main topic of this
Chapter.

Why bother with Monte Carlo simulation when there are so many
mean-field techniques, ranging from the simple Frumkin and Temkin
isotherms to sophisticated cluster-variation methods? This is a
natural question, and there are three compelling answers:
\begin{enumerate}
\item Mean-field methods are really not very accurate for most
      two-dimensional systems \cite{RIKV93D}. This is not to say that
      very useful results cannot be obtained by mean-field methods,
      for they certainly can, but simple mean-field methods ignore the
      spatial structure of the system. A nontrivial amount of work and
      attention goes into manually incorporating enough of this
      structure to give accurate results, especially in the vicinity
      of phase transitions. In contrast, Monte Carlo codes naturally
      take spatial fluctuations into account. Most of these codes are 
      quite simple to write and easily adapted to model different systems.
\item Monte Carlo methods are not limited to the study of equilibrium
      properties. Dynamic Monte Carlo simulations provide a tool to
      study time dependent phenomena in systems where spatial
      fluctuations may strongly modify the time evolution. Applied in
      this way, Monte Carlo simulation provides a more detailed
      supplement to standard rate-equation methods.
\item Computer technology has now evolved to a point where extensive
      simulations can be performed on work stations or personal
      computers that are well within the budget for most
      groups. Actually, several computers that are perfectly able to
      handle large Monte Carlo simulations may already be sitting in
      your laboratory! The competitive edge of methods that are
      analytically solvable or rely on numerical solution of
      relatively small sets of equations is much smaller than it was
      only a few years ago.
\end{enumerate}
The purpose of this Chapter is to discuss some of the basic aspects of
Monte Carlo simulation directly in the context of interfacial
electrochemistry. It is not intended to be a full textbook on Monte
Carlo simulation. There is no sense in duplicating the excellent
textbooks \cite{BIND92B,GOUL96} and large tutorial articles
\cite{BIND79,BIND92} that already exist.  Rather, we hope to
illustrate some of the details and pitfalls associated with
constructing and running Monte Carlo simulations of specific systems
of interest in surface electrochemistry.  

The outline of the rest of this Chapter is as follows. The lattice-gas
approach to modeling electrochemical adsorption is briefly introduced
in Sec.~\ref{sec:lg_def}. The process of comparing numerical and
experimental results to adjust the lattice-gas parameters is the
subject of Sec.~\ref{sec:fit}.  The basics of equilibrium and dynamic
Monte Carlo simulations as nonperturbative methods for analyzing
lattice-gas models at finite temperature are discussed in
Sec.~\ref{sec:mc}.  These simulations can be used to obtain numerical
results for such experimentally measurable quantities as adsorption
isotherms, voltammetric currents and charge densities, and images
obtained by either diffraction techniques or atomic-resolution
microscopies. In Sec.~\ref{sec:example}, we discuss three illustrative
examples drawn from our own research experience: the adsorption of
bromine onto Ag(100) \cite{OCKO97,KOPE97,WANG97}, the underpotential
deposition (UPD) of copper on Au(111) from a sulfate-containing
electrolyte \cite{BLUM96,HUCK90,BLUM91,HUCK91,ZHAN96,BLUM94A,BLUM94B},
and the electrosorption of urea on Pt(100) from an acid electrolyte
\cite{RIKV95,GAMB93B}. Particular emphasis is given to the
simulational aspects of these projects.  Since this is not a review
article, references to the original literature have been limited to
those we thought most illustrative for our purposes.

%
%
 
\section{Lattice-Gas Models of Chemisorption}
\label{sec:lg}

\subsection{Structure of the Models}
\label{sec:lg_def}

In lattice-gas models adsorption is only allowed on a discrete lattice
of adsorption sites.  The concentration of adsorbate at site $i$,
$c_i$, is $1$ if the site is occupied by the adsorbate and $0$
otherwise.
The energy of a
particular configuration of occupied sites, $c$, is given by the 
grand-canonical 
lattice-gas Hamiltonian
\begin{equation}
\label{eq:lg1}
{\cal H}(c)
         = - \bar{\mu} \sum_i c_i
           - \sum_n \left( \Phi^{(n)} \sum_{\langle ij \rangle}^{(n)} c_i c_j \right)
           - {\cal H}_3(c)
\;.
\end{equation}
The electrochemical potential of the adsorbate in solution is
$\bar{\mu}$, and the first term on the right-hand side is the
free energy of adsorption when adsorbate-adsorbate interactions are
ignored. The next term represents all of the two-body
adsorbate-adsorbate interactions. The index $n$ indicates the
separation between sites, {\it e.\/g.} $n=1$ indicates nearest
neighbors, while $\sum_{\langle ij \rangle}^{(n)}$ indicates the sum
over all pairs of neighbors of rank $n$. The interaction energy for
two adsorbate particles occupying $n$-th nearest neighbor sites is
$-\Phi^{(n)}$. Interactions between three or more adsorbed particles
are possible \cite{EINS91,EINS95} --- here they are symbolically
represented as the term ${\cal H}_3$.  The sign convention is such
that $\bar{\mu}$$>$0 denotes a tendency for adsorption in the absence
of lateral interactions, and $\Phi^{(n)}$$>$0 denotes an effective
attraction. Models of particular electrochemical systems differ in
their binding-site geometries and the values and ranges of the lateral
interactions.

The connection of the theoretical lattice-gas model to the chemical
and electrochemical variables that characterize the experimental
system is made through the electrochemical potential. The
electrochemical potential of ${\rm X}$ is related to its bulk activity
$[{\rm X}]$ and the electrode potential $E$ as
\begin{equation}
\label{eq:echempot}
\bar{\mu} = {\mu}^0 
+ RT \ln {[\rm X] \over [\rm X]^0} - zFE \;,
\end{equation}
where $R$ is the molar gas constant, $T$ is the absolute temperature,
$F$ is Faraday's constant, and $z$ is the effective electrovalence of
X. The quantities superscripted with a 0 are reference values which
contain the local binding energies. Here we treat them as adjustable 
model parameters. 

The thermodynamic density conjugate to the electrochemical potential is
the surface coverage
\begin{equation}
\label{eq:coverage}
\Theta = N^{-1}\sum_i c_i \;,
\end{equation}
where $N$ is the total number of adsorption sites.  With the sign
convention that oxidation currents are positive, the adsorbed charge
per adsorption site is
\begin{equation} 
\label{eq:charge}
q = -ez \Theta \;,
\end{equation}
where $e$ is the elementary charge unit.

Many interesting systems involve the cooperative chemisorption of two
different species, say ${\rm A}$ and ${\rm B}$, onto the surface. In
this case there is a concentration for each species at site $i$,
$c^{\rm A}_i$ and $c^{\rm B}_i$, with simultaneous adsorption of both
species at one site forbidden \cite{LEE79,HUCK84}. 
The lattice-gas Hamiltonian for such a
two-component system can be written as \cite{RIKV88B} 
\begin{eqnarray}
\label{eq:lg2}
\nonumber
{\cal H}\left(c\right) & = & -\bar{\mu}_A \sum_i c^{\rm A}_i
                              -\bar{\mu}_B \sum_i c^{\rm B}_i \\
& &
 - \sum_n \left[ 
   \Phi^{(n)}_{\rm AA} \sum_{\left<ij\right>}^{(n)} c^{\rm A}_i c^{\rm A}_j
 + \Phi^{(n)}_{\rm BB} \sum_{\left<ij\right>}^{(n)} c^{\rm B}_i c^{\rm B}_j
 + \Phi^{(n)}_{\rm AB} \sum_{\left<ij\right>}^{(n)}
   \left(  c^{\rm A}_i c^{\rm B}_j +  c^{\rm B}_i c^{\rm A}_j \right )
   \right]
 - {\cal H}_3
\;.
\end{eqnarray}
The total configuration of the system, which depends on both 
the configurations of A and B particles, $c^{\rm
A}$ and $c^{\rm B}$, is still represented as $c$.  The interactions
$\Phi^{(n)}_{\rm XY}$ (where ${\rm X}$ and ${\rm Y}$ can be ${\rm A}$
or ${\rm B}$) now depend on the adsorbed species as well as the
distance between sites. The higher-order interactions, ${\cal H}_3$,
may also involve combinations of ${\rm A}$ and ${\rm B}$.

Each component has its own electrochemical potential $\bar{\mu}_{\rm
X}$, surface coverage $\Theta_{\rm X}$, and effective electrovalence
$z_{\rm X}$. When the temperature is fixed, a phase diagram can be
constructed in the $(\bar{\mu}_{\rm A},\bar{\mu}_{\rm B})$ plane.  For
a specific chemical system with fixed bulk activities $[{\rm A}]$ and
$[{\rm B}]$, Eq.~(\ref{eq:echempot}) for $\bar{\mu}_{\rm A}$ and
$\bar{\mu}_{\rm B}$ form the parametric representation of a line in
the phase diagram. The line has slope $z_{\rm A}/z_{\rm B}$, and the
electrode potential $E$ is the parameter that determines the location
along this {\em isotherm\/}.  

Since the entropy vanishes at zero temperature, the $T=0$ phase
diagram, or ground-state diagram, can be calculated analytically using
Eq.~(\ref{eq:lg2}).  In principle this can be done directly, but in
practice another approach is often used.  For a given point in the
phase diagram, the ground state corresponds to the phase with the
lowest energy. A list of the conceivable ground states compatible with
the lattice symmetry \cite{SCHI81} and their energies is
generated. Then a ground-state diagram can be constructed
automatically by finding the lowest-energy phase directly while
scanning in the $\bar{\mu}_{\rm A}$ and $\bar{\mu}_{\rm B}$
directions. 

In contrast, the finite-temperature phase diagram for most systems can
only be mapped out using numerical methods, such as the Monte Carlo
simulations that are the main subject of this Chapter.  Since the
number of possible ground states for systems with longer-range
interactions can be very large, it is in practice advisable to check
the ground-state calculations and Monte Carlo simulations at low
temperature against each other for consistency.

When two types of ions adsorb onto the surface, the adsorbed charge
density is
\begin{equation}
q = -e \left( z_{\rm A}\Theta_{\rm A} + z_{\rm B}\Theta_{\rm B} \right)
\;.
\end{equation}
In the absence of diffusion and in the limit of vanishing potential
sweep rate \cite{BARD80}, the voltammetric current can be estimated in
terms of the equilibrium lattice-gas response functions
$\partial\Theta_{\rm X}/\partial\bar{\mu}_{\rm Y}$.
Specifically, using the chain rule this quasi-equilibrium current 
is obtained as 
\begin{equation}
\label{eq:idef}
i = \frac{{\rm d}q}{{\rm d}t} = 
  \frac{{\rm d}q}{{\rm d}E} \frac{{\rm d}E}{{\rm d}t}
\;. 
\end{equation}
The total differential of the charge density is
\begin{equation}
\label{eq:dq}
{\rm d}q = -e 
\left[ 
\left(
   z_{\rm A}\frac{\partial \Theta_{\rm A}}{\partial \bar{\mu}_{\rm A}}
 + z_{\rm B}\frac{\partial \Theta_{\rm B}}{\partial \bar{\mu}_{\rm A}} 
\right) {\rm d} \bar{\mu}_{\rm A}
+
\left(
   z_{\rm A}\frac{\partial \Theta_{\rm A}}{\partial \bar{\mu}_{\rm B}}
 + z_{\rm B}\frac{\partial \Theta_{\rm B}}{\partial \bar{\mu}_{\rm B}} 
\right) {\rm d} \bar{\mu}_{\rm B}
\right]
\;,
\end{equation}
which can be simplified with the Maxwell relation $\partial\Theta_{\rm
A}/\partial\bar{\mu}_{\rm B}= \partial\Theta_{\rm
B}/\partial\bar{\mu}_{\rm A}$.  For any particular electrochemical
experiment, ${\rm d}\bar{\mu}_{\rm X}=-z_{\rm X}F{\rm d}E.$
Substituting this into Eq.~(\ref{eq:dq}) and that result into
Eq.~(\ref{eq:idef}), the quasi-equilibrium current becomes
\begin{equation}
i = e F
\left[ 
   z^2_{\rm A}\frac{\partial\Theta_{\rm A}}{\partial\bar{\mu}_{\rm A}}
 + 2z_{\rm A}z_{\rm B}\frac{\partial\Theta_{\rm B}}{\partial\bar{\mu}_{\rm A}}
 + z^2_{\rm B}\frac{\partial\Theta_{\rm B}}{\partial\bar{\mu}_{\rm B}}
\right]
\frac{{\rm d}E}{{\rm d}t} \;.
\label{eq2b}
\end{equation}
The response function can, in turn, be estimated from the equilibrium
partition function using standard statistical-mechanics arguments, 
as described in the next section.

Coverages, charges, and currents are given here per adsorption site. 
They can be converted to the experimentally measured units of 
inverse surface area through division by the area per adsorption site. 

%
%

\subsection{Model Development and Refinement}
\label{sec:fit}

Accurate theoretical description of lateral interactions in an
adsorbed layer must consider the complete substrate-adlayer-solution
system. For instance, the geometry of adlayers observed {\em in situ}
for CO on Pt(111) are different for ultra-high vacuum, high-pressure
gas, and electrochemical environments \cite{JENS98}. In addition,
adsorption or solvation at one point on the surface can affect
adsorption at another point through deformation of the substrate or
distortion of the substrate electron structure.  Because of the
complicated nature of these systems, deriving accurate {\em ab initio}
estimates of interaction energies in electrochemical adsorption
problems is presently not feasible.  A practical alternative is to
treat the interactions as adjustable parameters. Because of the large
number of parameters and the time-consuming simulations, parameter
estimation by a formal optimization procedure must often be abandoned
in favor of adjusting the parameters manually to incorporate
experimental results and other chemical information efficiently.  This
method also has its difficulties. In particular, all of the potential
consequences of adjusting a parameter may not be easily foreseen, and
often several parameters must be adjusted in concert to change a
single aspect of the model. In addition, there is no guarantee that
the observations from two radically different sets of interactions are
measurably different in the region explored by the simulations.
Nevertheless, the encouraging results of previous lattice-gas studies
of electrochemical systems
\cite{BLUM90,BLUM96,RIKV91B,RIKV95,HUCK90,BLUM91,HUCK91,ZHAN96,BLUM94A,BLUM94B,GAMB93B,RIKV88B,RIKV91A,COLL89}
indicate that when proper attention is paid to including all available
experimental information consistently, this approach has considerable
predictive power. Furthermore, as effective interactions obtained by
first-principles calculations become available in the future, results
from lattice-gas models will provide crucial information for testing
the consistency of such first-principles interactions with
experimentally observed thermodynamic and structural information.

The modeling strategy starts by using theoretical and experimental
knowledge about the substrate structure and the shapes and sizes of
the adsorbate particles to formulate a specific lattice-gas model.
The coverages and adlayer structure of different adsorbate phases can
be determined from microscopy, scattering, radiochemical
experiments, or standard electrochemical methods.  
Group-theoretical ground-state calculations
\cite{SCHI81} are then performed to determine a minimal set of
effective interactions consistent with the observed adsorbate phases
\cite{RIKV91B,RIKV88B,RIKV91A,COLL89}. A ground-state diagram
is obtained by pairwise equating the ground-state energies of the
different phases, as obtained from the lattice-gas Hamiltonian, {\em
e.g.} Eq.~(\ref{eq:lg1}) or Eq.~(\ref{eq:lg2}). A typical example,
corresponding to the model discussed in Sec.~\ref{sec:cu}, is shown in
Fig.~\ref{fig:gstate}.

The finite-temperature properties of the lattice gas, calculated by
Monte Carlo or other methods,
\cite{BLUM96,RIKV91B,RIKV95,ZHAN96,GAMB93B,RIKV88B,RIKV91A,COLL89}
can be compared to available experiments and used to refine
the effective interactions. The process should be iterated until
satisfactory agreement between model and experiment is achieved. New
experiments may be required to clarify the properties of the chemical
system and refine the theoretical model.

%
%

\section{Monte Carlo Simulation}
\label{sec:mc}

\subsection{Monte Carlo Sampling of Equilibrium Configurations}
\label{sec:sample}

In statistical mechanics the properties of a system in equilibrium,
such as the coverages and response functions 
mentioned in the previous section, are
calculated from the partition function
\begin{equation}
\label{eq:partition}
Z = \sum_{\left\{c\right\}} \exp{\big[-{\cal H}(c)/RT\big]}
\;,
\end{equation}
where $\sum_{\{c\}}$ is over all possible configurations of the
adsorbates.  The expectation value of a quantity $Y$ at equilibrium is
the weighted sum
\begin{equation}
\label{eq:expectation}
\left<Y\right>=\frac{1}{Z}\sum_{\left\{c\right\}} Y(c) \exp{\big[-{\cal H}(c)/RT\big]}
\;,
\end{equation}
where the weight is a Boltzmann factor.  For most realistic models it
is hard to calculate, or even estimate, Eq.~(\ref{eq:partition}) or
(\ref{eq:expectation}) analytically.  On the other hand, Monte Carlo
simulation methods can be used to estimate the properties of even
quite complicated models numerically \cite{BIND92B}.

The reason for our advocacy of the nonperturbative numerical Monte
Carlo approach is that such calculations are much more accurate for
two-dimensional systems than even quite sophisticated mean-field
approximations \cite{RIKV93D}. From Monte Carlo simulations one can
obtain thermodynamic and structural information \cite{BIND92B,BIND92},
using modest computational resources and programming effort.  In
combination with finite-size scaling analysis \cite{BIND92}, Monte
Carlo methods have contributed significantly to the theoretical
understanding of fluctuations and ordering in adsorbate systems.
Monte Carlo codes also have the advantage that they are relatively
easy to modify to accommodate changes in lattice structure and/or
interaction geometries and ranges.  This is useful when one wants to
modify the model to improve the agreement with experiments, or
refurbish an existing code to model a different system.

One Monte Carlo method for estimating Eq.~(\ref{eq:expectation}) is to
evaluate the sum directly from randomly generated adsorbate
configurations. For most problems where temperature is important, this
method is inefficient since only the most energetically favorable
configurations make significant contributions to the sum. A much more
efficient method, called {\em importance sampling}, samples more
frequently from the energetically favorable configurations. Usually
this is done by using the Boltzmann weight, the probability of an
equilibrium system being in a particular state, to determine the sampling
probability of that state. In addition, one can use the idea of {\em
detailed balance}, that the rate of going from state ${c}$ to state
${c'}$ in equilibrium is the same as the rate of going in the opposite
direction. If ${\cal N}({c})$ is the density of states for
configuration ${c}$ and the probability of going from configuration
${c}$ to configuration ${c'}$ during a certain period of time is
${\cal R}({c} \rightarrow {c'})$, the transition rate is ${\cal
N}({c}){\cal R}({c} \rightarrow {c'})$. Detailed balance implies that
\begin{equation}
\label{eq:detail1}
\frac{ {\cal N}({c}) }{ {\cal N}({c'})} =
\frac{ {\cal R}({c'} \rightarrow {c}) }{ {\cal R}({c} 
\rightarrow {c'}) }
\;.
\end{equation}
To sample configurations according to the Boltzmann weights, the
transition probabilities need to obey
\begin{equation}
\label{eq:detail2}
\frac{ {\cal R}({c'} \rightarrow {c}) }{ {\cal R}({c}
\rightarrow {c'}) } = \exp{ \left( -\frac{ {\cal H}({c}) -
{\cal H}({c'})}{RT} \right) } 
\;.
\end{equation}

An efficient Monte Carlo simulation method for sampling the
equilibrium ensemble of a lattice gas is obtained by considering a
long series of small, random changes in the concentration that obey
detailed balance.  Before considering particular algorithms, it is
necessary to factor the transition probability into the probability of
attempting a move, ${\cal A}({c} \rightarrow {c'})$, and the
probability of accepting the move ${\cal P}({c} \rightarrow {c'})$, 
{\it i.\/e.} ${\cal R}({c} \rightarrow {c'}) = 
{\cal A}({c} \rightarrow {c'}) {\cal P}({c} \rightarrow {c'})$. 

A commonly used and intuitively appealing Monte Carlo algorithm is 
the Metropolis algorithm. Given a particular
configuration, all possible changes to the configuration are given the
same attempt probability ${\cal A}$. The attempted move is then accepted
with probability
\begin{equation}
\label{eq:Metropolis}
{\cal P}({c} \rightarrow {c'}) = 
{\rm min}
  \left(
  \exp{\left[-\frac{{\cal H}({c'})-{\cal H}({c})}{RT} \right]}, 
  1
  \right)
\;,
\end{equation}  
which can be verified to obey the detailed-balance condition,
Eq.~(\ref{eq:detail2}), for all cases. A similar algorithm, known
by the name Glauber \cite{GLAU63}, also obeys detailed balance. It
accepts the move with probability
\begin{equation}
{\cal P}({c} \rightarrow{c'}) = 
  \frac{1}{1+\exp{\left[\left({\cal H}({c'})-{\cal H}({c})\right)/RT\right]}}
\;.
\end{equation}
The two algorithms become identical as $T \rightarrow 0$, except for
${\cal H}({c'})-{\cal H}({c})=0$.

Another importance-sampling algorithm is useful when more than two
configurations are possible for the portion of the surface being
updated. For one move in this {\em heat bath} algorithm,
the Boltzmann factor for every possible configuration is calculated,
and these weights are used to randomly choose the final configuration.
The new configuration is accepted automatically, with the initial
configuration as one of the possible choices.  In the case only two
configurations are possible, the heat-bath algorithm reduces to the
Glauber algorithm \cite{GLAU63}. That this algorithm satisfies
detailed balance can be easily verified.

Commonly, Monte Carlo updates change only one adsorption
site. However, in many applications equilibration can be slow, either
because of the large fluctuations that occur near second-order phase
transitions (critical slowing-down \cite{BIND92B,BIND92}) or because of large
local energy barriers. In such cases {\em cluster update} algorithms
\cite{CHAY97} that involve several related sites in one Monte Carlo
update can be useful. A specific example of a heat-bath cluster
algorithm designed to overcome slowing-down due to high local barriers
against single-particle moves is discussed in Sec.~\ref{sec:urea}.

When importance sampling is used to incorporate the weighting, many
thermodynamic averages can be estimated from the moments of the sampled
quantities. For example, the $q$-th moment of the number of adsorbed
particles of species ${\rm X}$ is
\begin{equation}
M_q^{\rm X} = s^{-1}\sum_{j=1}^{s} \left( \sum_{i=1}^N c_i^{\rm X}  
\right)_j^q
\;,
\end{equation}
where $s$ is the number of samples taken during the
simulation. The coverage is then just
\begin{equation} 
\Theta_{\rm X} = 
\frac{1}{N}\frac{\partial\ln{Z}}{\partial(\bar{\mu}_{\rm X}/RT)}
       = \frac{M_1^{\rm X}}{N}
\;,
\end{equation}
with $N$ the number of adsorption sites. The response functions,
$\partial\Theta_{\rm X}/\partial\bar{\mu}_{\rm X}$, which are related to the
quasi-equilibrium CV current (see Sec.~\ref{sec:lg}), are 
\cite{varNote}
\begin{eqnarray}
\label{eq:chi}
\frac{\partial\Theta_{\rm X}}{\partial\bar{\mu}_{\rm X}}
     & = & \frac{1}{RT} 
                 \frac{\partial^2\ln{Z}}{\partial(\bar{\mu}_{\rm X}/RT)^2} 
                 \nonumber\\
     & = & \frac{s}{s-1}\frac{M_2^{\rm X} - (M_1^{\rm X})^2}{N\, RT}
\;.
\end{eqnarray}
The response function is proportional to the variance of the number of
adsorbed ${\rm X}$ particles, which corresponds to the fluctuations in
the coverage. Equation (\ref{eq:chi}) is an example of the class of
important results in statistical mechanics, called
fluctuation-dissipation theorems \cite{PATHRIA}. In our practical context
it provides a convenient way to calculate the response functions
without numerical differentiation. The cross-response function,
$\partial\Theta_{\rm X}/\partial\bar{\mu}_{\rm Y}$, is calculated
entirely analogously, using the covariance of ${\rm X}$ and ${\rm Y}$.

In thermodynamic regions far from any phase transitions, the
fluctuations in the numbers of adsorbate particles have a small
correlation length.  The surface can therefore be divided into
independent pieces, and the fluctuations in adparticle numbers are
proportional to the surface area.  As a consequence the {\em
relative\/} fluctuations, which are proportional to $\sqrt{\partial
\Theta_{\rm X} / \partial \bar{\mu}_{\rm X} } / \Theta_{\rm X}$ vanish
with system size as $N^{-1/2}$.  Near a second-order phase transition
on the other hand, the correlation length becomes comparable to the
system size, and the fluctuations and response functions depend on $N$
as power laws.  The power-law exponents depend on the nature of the
phase transition, and can be studied by finite-size scaling techniques
\cite{BIND92,ZHAN96,GOLD92}.

The actual execution of a Monte Carlo simulation involves several
steps. An initial configuration is constructed, followed by a
sufficient number of Monte Carlo updates to allow it to reach
equilibrium. A good habit is to construct the initial
configuration using different phases for the right and left halves of
the simulated system. This reduces the probability of relaxing to a
long-lived metastable phase rather than true equilibrium. Also,
several quantities should be sampled periodically during the
equilibration process to ensure that the system has indeed reached
thermodynamic equilibrium before equilibrium sampling begins.
Equilibrium samples should be spaced far enough apart to be
statistically independent. This is normally not a problem in
parameter regions far from any phase transitions. Near phase
transitions, however, critical slowing-down leads to very strong
correlations between subsequent configurations and requires the
sampling to be performed at longer intervals.  While sampling with
intervals that are too short leads one to underestimate fluctuations
and thereby response functions, sampling too infrequently is a waste of
computer time. Optimal sampling can be achieved by calculating time
correlation functions \cite{BIND92B} of the quantities being sampled
and adjusting the interval accordingly. To improve accuracy, averages
should also be performed over several trials, each consisting of
equilibration followed by sampling.

Finally, it must be noted that there is a difference between a single
attempted move, often called a {\em Monte Carlo step} (mcs), and the
more frequently encountered {\em Monte Carlo step per site}
(MCSS). The later is useful because it is independent of the number of
lattice sites and represents the smallest unit of updates over which
the lattice as a whole can relax. These units should be distinguished
clearly when reporting the results of Monte Carlo simulations.

\subsection{Dynamic Monte Carlo Simulations}

\label{sec:dynamic}

In the previous Subsection we discussed the basic interpretation of
Monte Carlo simulation as a method to measure the equilibrium
partition function and moments of quantities of interest. When it is
used in this way, the simulation says nothing about the time
dependence of the system. Consequently, one can use {\em any\/}
algorithm that speeds up equilibration and reduces the correlations
between subsequent configurations.  The only requirement is that it
must obey detailed balance in order to ensure that thermodynamic
equilibrium is indeed reached and importance sampling can be used.

A very different application of Monte Carlo simulation results when
the number of MCSS is viewed as {\it time\/}. Then the simulation can
be viewed as a ``movie'' of the dynamic process.  Several examples of
such movies can be found at the authors' Web site \cite{URL}.

It is clear that a particular Monte Carlo algorithm does not
necessarily say anything about the time evolution of the corresponding
chemical or physical system unless it is specifically constructed to
do so. It is possible to construct a Monte Carlo simulation that evolves
like the physical system, at least for dynamics that can be reasonably
modeled as a stochastic process. This is the case when the dynamics
are dominated by thermally activated processes such as adsorption,
desorption, and diffusion. Here we describe the recipe for
transforming a lattice-gas Hamiltonian, hopefully determined to
reasonable accuracy by equilibrium studies, into such a dynamical
model.

Essentially, every mcs should correspond to a thermally activated
transition between states separated by energy barriers, with the
transition rates described by Arrhenius rates involving the barrier
heights. These barrier heights can be constructed by a method
essentially equivalent to the Butler-Volmer approximation
\cite{BARD80}. This is done in the framework of activated-complex
theory, or transition-state theory, with the initial and final states
described by the lattice gas. The activated complex cannot be
represented by the lattice gas, but a free energy can be assigned to
it using the model we now describe.

A schematic free-energy surface is presented in Fig.~\ref{fig:Barrier}
for several possible lattice-gas configurations, $^kc$. First consider
the two states with free energy ${\cal H}( ^1c)={\cal H}({^2c})=0$
(depicted as solid line segments). The horizontal line segments
correspond to the lattice-gas states, which are locally stable. The
free energy is taken to increase linearly as the system moves toward
the activated complex, which occurs where the two free-energy surfaces
intersect. This linear assumption for the free-energy increase leads
directly to the Butler-Volmer equation \cite{BARD80}. The height of
the free-energy barrier associated with this case, $\Delta$, is a
parameter which can be adjusted to describe particular processes.

In general the slope of the free energy can also be adjusted to
describe specific chemical systems. If the two slopes are described by
the angles $\theta$ and $\phi$, as shown in Fig.~\ref{fig:Barrier},
then changes in the activated-complex free energy associated with
${\cal H}(c) \neq 0$, can be found in terms of the {\em symmetry
factor}
\begin{equation}
\alpha = \frac{\tan{\theta}}{\tan{\theta}+\tan{\phi}}
\;.
\end{equation}
For a transition from the ${\cal H}({^1c})=0$ state to the
${\cal H}({^4c}) \neq 0$ state, the free energy of the activated complex
is
\begin{equation}
{\cal H}^*\left({^1c},{^4c}\right) = 
\Delta + \left(1-\alpha\right) {\cal H}\left({^4c}\right)
\;.
\end{equation}
When $\alpha \neq 1/2$, 
Fig.~\ref{fig:Barrier} is not symmetric about the activated
complex. So if ${\cal H}({^3c}) \neq 0$ while ${\cal H}({^2c})=0$,
the activated-complex free energy is
\begin{equation}
{\cal H}^*\left({^3c},{^2c}\right) = 
\Delta + \alpha {\cal H}\left({^3c}\right)
\;,
\end{equation}
where the symmetry when $\alpha=1/2$ is obvious. For the most general
case with ${\cal H}({^3c})$ and ${\cal H}({^4c})$, the free energy of the
activated complex is
\begin{equation}
{\cal H}^*\left({^3c},{^4c}\right) = \Delta +
\left(1-\alpha\right) {\cal H}\left({^3c}\right)+
\alpha {\cal H}\left({^4c}\right)
\;.
\end{equation}
This free energy is unaltered by reflection around the transition
state. In addition, the energy barriers ${\cal H}^* - {\cal H}\left(
^3c\right)$ and ${\cal H}^* - {\cal H}\left({^4c}\right)$ are not
altered by shifting the zero of the free-energy scale.

The symmetric case, $\alpha=1/2$, gives the convenient relation 
\begin{equation}
\label{eq:barrier}
{\cal H}^*\left({c},{c'}\right) = \Delta +
 \frac{1}{2}\Big[{\cal H}\left({c}\right)+{\cal H}\left({c'}\right)\Big]
\;.
\end{equation}
In this case the activated-complex free energy is found by adding a constant
barrier to the arithmetic mean of the free energies of the lattice-gas
states \cite{VATT96}. Once a consistent barrier scheme has been
identified, a number of reasonable transition rates can be
constructed, subject only to the requirement of detailed balance. The
simplest rate is probably the one given by
\begin{equation}
\label{eq:gregRate}
R\left({c} \rightarrow{c'}\right) = \nu_0 \exp{\left(-\frac{\Delta}{RT}\right)}
\exp{\left(-\frac{{\cal H}\left({c'}\right)-{\cal H}\left({c}\right)}
{2RT}\right)}
\;,
\end{equation}
where $\nu_0$ is the attempt frequency, typically on the order of a
phonon frequency ($10^9$ -- $10^{13}$~Hz). Another useful choice is
the Transition Dynamics Algorithm \cite{VATT92},
\begin{equation}
{\cal R}(c \rightarrow c') =
\frac{\nu_0}
{\left[1+e^{\left({\cal H}^*(c,c')-{\cal H}(c)\right)/RT}\right]
 \left[1+e^{\left({\cal H}(c')-{\cal H}^*(c,c')\right)/RT}\right]
}
\;,
\end{equation}
which is simply the product of the Glauber transition rates from $c$
to the activated complex, and from the activated complex to $c'$.

The single-ion processes included in the dynamical model discussed
here are diffusion to nearest-neighbor sites and adsorption/desorption
at single sites. We assume a single value for the angle in
Fig.~\ref{fig:Barrier} for all configurations with the moving ion near
the surface, which implies $\alpha=1/2$ for diffusion. In general, the
angle associated with an ion far from the surface has a different
value, but we make the reasonable \cite{BARD80} choice $\alpha=1/2$ so
Eq.~(\ref{eq:barrier}) can be used for all microscopic processes.  On
the other hand, we use a different barrier, $\Delta_{\rm X}^{\rm P}$,
for each combination of ion and process ${\rm P}$ ( ${\rm P}={\rm a}$
for adsorption/desorption and ${\rm P}={\rm d}$ for diffusion). The
values of the $\Delta_{\rm X}^{\rm P}$ determine the relative weights
among the different microscopic processes.  Reasonable agreement with
experimental time scales would indicate that the combination of attempt
frequencies and barrier heights chosen is realistic.

\subsection{The $n$-fold Way}

In many situations of experimental interest, a direct implementation
of an importance-sampling Monte Carlo algorithm with Butler-Volmer
transition rates would require impossibly long simulation times. The
basic reason for this is that only a very small fraction of attempted
Monte Carlo updates may actually result in a change of the current
configuration. Since we are now interested in the dynamical details,
we cannot solve this problem by constructing a cluster algorithm,
since this would generally change the dynamics completely.  What is
needed is an algorithm which can simply reduce the computer time
needed to perform the large number of unsuccessful attempts that would
be made between each successful update while remaining faithful to the
original dynamics.  A class of algorithms that perform this feat are
known as {\it rejection-free algorithms\/} \cite{BORT75,NOVO95}.  Many
such algorithms have been invented and re-invented over the years (see
Ref.~\cite{NOVO95}), they are all essentially equivalent. In these
algorithms the acceptance probability is unity, but the probability of
attempting a move is based on its Boltzmann weight. They can also be
thought of as a Metropolis or heat-bath simulation with the number of
rejected moves before the next successful move calculated
analytically. The $n$-fold way \cite{BORT75}, introduced by Bortz,
Kalos, and Lebowitz, is one of the best known and earliest of these
methods.

Rejection-free algorithms are implemented by keeping a list of all the
possible moves. Each move is weighted by its rate ${\cal R}$, so that
making a random choice from the list is biased towards faster
processes. The stochastic nature of real processes is captured by
updating the simulation clock after each move by a random number that
represents the time elapsed since the last move. For the current
configuration, the overall rate for the system to have any move
accepted is the sum of all the individual move rates, ${\cal R}_{\rm
c}$. Since the probability of escaping from the current configuration
has a constant rate, the elapsed time until the accepted change has an
exponential distribution, and can be found using
\begin{equation}
\Delta t = -{\cal R}_{\rm c}^{-1} \ln{(r)}
\;,
\end{equation}
where $r$ is a random number uniformly distributed on $(0,1)$.
Here $\Delta t$ is a continuous time in units of mcs.

The weighted list can be visualized as a line segment whose length
equals the sum of all the weights. Then, choosing from the
weighted list is accomplished by picking a random point along the line
segment, {\em i.e.} picking a number on the interval [0,1) and
multiplying by the segment length. Determining which choice this
corresponds to involves searching the list of process rates. This
searching can become an important bottle neck in the implementation of
$n$-fold way algorithms, but the effect can be reduced by using a
binary tree to search for the choice from the weighted list. Building
a binary tree of partial sums (of the rates) to speed up the search is
not difficult, and a great deal of help is available from many
introductory texts on data structures \cite{AHO}. The rates make up
the leaves at the bottom of the tree, the value of each node is the
sum of all the leaves below it. As the simulation proceeds, the values
of the leaves change and the nodes must be updated. Two approaches can
be taken. The simplest is to move up the tree from the changed leaf,
explicitly adding the values of the children to find the value of the
node. A faster approach is to calculate the change in the leaf value,
and add this to all the nodes above the leaf. Unfortunately, with
this faster approach truncation errors will eventually cause the nodes
to not accurately reflect the values of the leaves. Practically this
method can be used, but the entire tree should be recalculated every
hundred thousand MCSS or so.

%
%

\section{Specific Examples}
\label{sec:example}

\subsection{Br on Ag(100)}
\label{sec:br}

Experimental CVs of bromine adsorption onto silver (100)
single-crystal surfaces display a broad peak followed by a stronger,
sharper peak at more positive potentials \cite{OCKO97,VALE78}. {\em In
situ} surface x-ray scattering shows that the sharp peak corresponds
to a second-order, {\em i.\/e.} continuous, phase transition from a
disordered low-coverage phase at negative potentials to an ordered
$c(2\times2)$ phase at more positive potentials \cite{OCKO97}. Since
the adsorption sites for the bromine are the four-fold hollow sites of
the square Ag(100) surface \cite{KOPE97,WANG97}, it is natural to
model this system with a lattice-gas Hamiltonian of the form given by
Eq.~(\ref{eq:lg1}), using a square lattice of adsorption sites.

The ordered phase corresponds to a maximum bromine coverage of
$\Theta=0.5$.  This is related to the adsorbed bromine being bigger than the
Ag(100) lattice spacing, such that adsorption in nearest-neighbor
hollow sites is very unfavorable energetically.  This packing feature
can be modeled as an infinitely strong repulsion between
nearest-neighbor bromine, $\Phi^{(1)} = - \infty$. 
Relaxing the nearest-neighbor repulsion to
finite, but large, values does not significantly improve the agreement
with experiment \cite{KOPE97}. For separations larger than this,
pair interaction energies are modeled by unscreened dipole-dipole
interactions:
\begin{equation}
\label{eq:BrPhi}
\Phi^{(n)} = \frac{2^{3/2}\Phi^{(2)}}{r{(n)}^3}
\;,
\end{equation}
where $r{(n)}$ is the distance between sites, measured in units of the
lattice constant. This interaction falls off reasonably rapidly with
distance, and only interactions for $n \le 13$ were considered in the
simulations discussed here. This corresponds to approximately $87$\%
of the total dipole-interaction energy. Other lattice-gas simulations
\cite{KOPE97} used screened dipole-dipole interactions, but the effect
of screening is relatively small for Br$^-$ at lengths less than the
truncation used here.

If $\Phi^{(2)}=0$, the only interaction is the infinite
nearest-neighbor repulsion, and the lattice-gas model is just a
hard-square model \cite{BAXT80,GIAC95}. The hard-square model has a
critical point at $\bar{\mu} \approx 1.334 RT$, which corresponds to
the onset of percolation \cite{GIAC95} at a critical coverage
$\Theta_c \approx 0.37$ \cite{RACZ80,KINZ81}. Below this concentration
a disordered gas of adsorbate particles forms, but above $\Theta_c$ an
ordered $c(2\times2)$ phase with coverage near $1/2$ forms. Typical
configurations of the two phases are illustrated in
Fig.~\ref{fig:BrSnap}.  Numerical studies of the hard-square model
with repulsive next-nearest neighbor interactions \cite{KINZ81,BIND80}
indicate that a $\Theta=1/4$ phase (which is not seen experimentally
in the system studied here) also becomes important for strongly
repulsive $\Phi^{(2)}$.  This constrains the repulsive interactions in
the lattice-gas model. If the interactions are too strong, a plateau
in the isotherm occurs around $\Theta=1/4$, in disagreement with
experiments.

To generate a numerical adsorption isotherm, the bromine coverage at
equilibrium was measured from Monte Carlo simulations at a discrete
set of electrochemical potentials using a simple Metropolis algorithm.
Since no finite-size scaling analysis was planned, only one lattice
size was used: a $32\!\times\!32$ system with periodic boundary
conditions. The size and boundary conditions used illustrate general
rules for simulation of lattice-gas models:
\begin{itemize} 
\item The system size must be chosen as a multiple of the
      periodicities of all important ordered phases. In the present 
      case this is easy, since the $c(2\!\times\!2)$ phase has 
      period two in both directions.  
\item Unless there is a specific reason to do otherwise, one
      eliminates edge and corner effects by using periodic boundary
      conditions. Geometrically, this can be visualized as wrapping
      the system onto a torus. The system size should be larger than
      the longest-ranged interaction to avoid wrapping all the way
      around.
\item Near critical points, fluctuations are correlated over lengths
      comparable to the system size (see Sec.~\ref{sec:mc}). Accurate 
      simulation near critical points therefore requires large systems.
\end{itemize}
Some sophistication is needed in the implementation to make the
simulation reasonably efficient. The long-range repulsion is truncated
at the $13$-th neighbor, a separation of $5$ lattice constants.  Due
to the long interaction range, a significant amount of computer time
is spent evaluating the energy change due to an attempted move. To
this end, a data structure tracking the number of $n$-th neighbor
bromines for each site is maintained and updated only after each
successful move. For this simulation the Metropolis acceptance rate is
around $10\%$, so this leads to a considerable speed-up. A more
efficient simulation using a heat-bath algorithm with cluster updates
is currently being investigated.

The experimental isotherm, determined by chronocoulometry
\cite{OCKO97,WANG97} for $0.1\,{\rm mM}$ Br, is shown in
Fig.~\ref{fig:BrIsotherm} (circles), along with two theoretical
results. The effective electrovalence, $z$, can be determined from the
experimental isotherms by assuming the coverage depends only on the
electrochemical potential and then fitting $z$ to isotherms for
different electrolyte concentrations using
Eq.~(\ref{eq:echempot}). Using the experimental results from
Ref.~\cite{WANG97} we find $z \approx -0.736$, which is in good
agreement with other estimates \cite{KOPE97,WANG97}.  With $z$ fixed
at this value and $\Phi^{(2)}$ taken as a fitting parameter, the best
fits for both the single-sublattice Frumkin isotherm \cite{WANG97}
(dotted line) and the full lattice-gas model evaluated by Monte Carlo
simulations give $\Phi^{(2)} \approx -24\,{\rm meV} \approx -2.3
\,{\rm kJ/mol}$.

Both theoretical isotherms agree qualitatively with the experimental
data, but there is a noticeable difference near $\Theta \approx 0.35$
where the lattice-gas model shows the singularity in the slope of the
isotherm associated with the phase transition.  The Frumkin isotherm
does {\it not\/} contain this singularity, but only a gradual
cross-over from low to high coverage. In this region, there is
considerably better agreement between the experimental and Monte Carlo
results. Reproducing this behavior is important, because the
singularity gives rise to a peak in the quasi-equilibrium CV (not
shown here) which corresponds to the peak seen in experiments.  Thus,
the Monte Carlo simulations provide a theoretical description of the
essential part of this electrochemical adsorption process that the
mean-field Frumkin isotherm cannot.

\subsection{Cu on Au(111) in the Presence of Sulfate}
\label{sec:cu}

In underpotential deposition (UPD), a monolayer or less of one metal
is electrochemically adsorbed onto another in a range of electrode
potentials more positive than those where bulk deposition occurs
\cite{BARD80}. The UPD of copper on Au(111) electrodes in
sulfate-containing electrolytes has been intensively studied, both
experimentally and theoretically. A discussion of the literature until
1995 is included in Ref.~\cite{ZHAN96}. The most striking feature in
CV experiments on this system is the appearance of two peaks,
separated by $\sim$ $50$-$150$ mV, upon addition of Cu$^{2+}$ ions to the
sulfuric-acid electrolyte \cite{SCHU76,DICK76,KOLB87,ZEI87}.  Typical
simulation CV profiles, estimated by both quasi-equilibrium and
dynamic methods, are shown in Fig.~\ref{fig:UPDcv}.  In the potential
range between the peaks, the adsorbate layer is believed to have a
($\sqrt3\times\sqrt3$) structure consisting of $2/3$ monolayer (ML)
copper and $1/3$ ML sulfate \cite{HUCK91}.  This structure, which is
illustrated in Fig.~\ref{fig:gstate}, is strongly
supported by {\em in situ} x-ray diffraction data \cite{TONE95}.

The two-component lattice-gas model used in references
\cite{ZHAN96,RIKV97A}, is a refinement of the original
Huckaby and Blum model for this system. It was extended with interactions of
intermediate range, similar to the extensions to the original model
discussed in Ref.~\cite{BLUM96}.  These models are based on the
assumption that the sulfate coordinates the triangular Au(111) surface
through three of its oxygen atoms, with the fourth S-O bond pointing
away from the surface.  This gives the sulfate a
triangular footprint with a O-O distance that reasonably matches
the lattice constant of the triangular Au(111) unit cell.  The copper
is assumed to compete for the same adsorption sites as the sulfate.
The configuration energies are given by Eq.~(\ref{eq:lg2}) with ${\rm
A}={\rm C}$ for copper and ${\rm B}={\rm S}$ for sulfate.

Just as for the Br/Ag(100) system discussed in Sec.~\ref{sec:br}, the
sulfate ion is sufficiently large that steric interactions prevent two
sulfates from binding simultaneously to nearest-neighbor sites.  This
is modeled as an infinite repulsion between nearest-neighbor
sulfates, $\Phi_{\rm SS}^{(1)} = - \infty$. Neglecting other
interactions, this corresponds to the hard-hexagon model
\cite{BLUM91,HUCK91,BLUM94A,BLUM94B} which can be solved analytically
\cite{BAXT80,BAXTER}. In the limit of strong sulfate adsorption,
$\bar{\mu}_{\rm S} \rightarrow + \infty$, the infinite
nearest-neighbor repulsion yields a 1/3 ML sulfate phase. Denoting a
phase with a $X\!\times\!Y$ unit cell as
($X\!\times\!Y$)$_{\Theta_{\rm C}}^{\Theta_{\rm S}}$, the highest
density sulfate-only phase is $(\sqrt{3}\!\times\!\sqrt{3})_0^{1/3}$.

The (rather complicated) complete ground-state diagram for a lattice
gas model of copper UPD is shown in Fig.~\ref{fig:gstate}.
Experimentally, the two CV peaks separate a copper monolayer, an
ordered $(\sqrt{3}\!\times\!\sqrt{3})_{2/3}^{1/3}$ phase, and a disordered
low-coverage $(1\!\times\!1)^0_0$ phase. In experiments involving
sulfate but no copper, the $(1\!\times\!1)^0_0$ and
$(\sqrt{3}\!\times\!\sqrt{7})_0^{1/5}$ phases are observed
\cite{MAGN90,EDEN94}.  The desire to reproduce this sulfate-only phase
largely dictated the values of the effective sulfate-sulfate
interactions used in the model \cite{ZHAN96}.  In the model, more
positive electrode potentials cause the sulfate to form its saturated
$(\sqrt{3}\!\times\!\sqrt{3})_0^{1/3}$ hard-hexagon phase. However, in
reality other chemical processes, such as surface oxidation, probably
occur before this phase is fully achieved.

The effective electrovalences, $z_{\rm S}$ and $z_{\rm C}$, must be
determined from experiments \cite{OMAR93,ZSHI94C,ZSHI95B}.  Zhang, {\em
et al.}\/, \cite{ZHAN96} obtained the values, $z_{\rm C} \approx+1.7$
and $z_{\rm S}\approx-1.1$, using data for the dependence of the
CV-peak separation on electrolyte concentration from Omar {\em et al.}
\cite{OMAR93}.  These values are consistent with independent estimates
\cite{ZSHI94C,ZSHI95B}.  As explained in Sec.~\ref{sec:lg},
constant-concentration isotherms are lines of slope $z_{\rm S}/z_{\rm
C}$ in the phase diagram. The isotherm used for this discussion is
given by the dashed line in Fig.~\ref{fig:gstate} for $100 \le E \le 300\,
{\rm mV}$ vs Ag/AgCl.

Starting from the negative end, we scan in the direction of positive
electrode potential (upper left to lower right in the figure). Near
the CV peak, at approximately 185~mV vs Ag/AgCl, the sulfate
begins to compete with copper for the gold surface sites, resulting in
a third of the copper desorbing into the bulk and being replaced by
sulfate.  Due to the strong effective attraction between the copper
and sulfate adparticles, a mixed
$(\sqrt{3}\!\times\!\sqrt{3})_{2/3}^{1/3}$ phase is formed (see
Fig.~\ref{fig:gstate}), which extends through the entire potential
region between the two CV peaks.  As the CV peak at approximately
250~mV is reached, most of the copper is desorbed within a narrow
potential range.  As it is thus deprived of the stabilizing influence
of the coadsorbed copper, the sulfate is partly desorbed, reducing
$\Theta_{\rm S}$ from 1/3 to approximately 0.05.

Both equilibrium and time-dependent Monte Carlo simulations have been
performed for this system, using the same model parameters as in
Ref.~\cite{ZHAN96}. The equilibrium simulations, which produced the
quasi-equilibrium CV currents shown in Fig.~\ref{fig:UPDcv}, were
performed with a heat-bath algorithm. To circumvent energy barriers
that reduce the acceptance rate for single-site updates, a
cluster-update algorithm treating two nearest-neighbor sites
simultaneously was employed.  The simulations were performed on a
rhombus-shaped triangular-lattice system of size $30 \times 30$ with
periodic boundary conditions. At each value of the electrode
potential, the system was equilibrated for between 50 and 500~MCSS,
after which sampling was performed every 5~MCSS for up to
5000~MCSS. Averaging over 20 trials was performed, except near the
sharp peak, where 215 trials were conducted.

Dynamic simulations were also performed on a $30 \times 30$ system,
using an $n$-fold way algorithm which includes adsorption, desorption,
and nearest-neighbor diffusion moves for both Cu and sulfate. We chose
$\nu_0=10^{10}\,{\rm s}^{-1}$ in Eq.~(\ref{eq:gregRate}) and used the
energy barriers $\Delta_{\rm S}^{\rm d}=\Delta_{\rm C}^{\rm
d}=35\,{\rm kJ/mol}$, $\Delta_{\rm S}^{\rm a}=55\,{\rm kJ/mol}$, and
$\Delta_{\rm C}^{\rm a}=50\,{\rm kJ/mol}$.  The result for a dynamic
simulation of a CV with scan rate ${\rm d}E/{\rm d}t=10\,{\rm mV/s}$
are also included in Fig.~\ref{fig:UPDcv}.  After initial
equilibration at a constant electrode potential and room temperature
for 1~sec, the dynamic simulation proceeds by changing $E$ in steps of
1~mV, simulating for a time determined by the scan rate, and then
sampling once. This process is continued until the end of the scan is
reached. The results from each potential scan are averaged over 75
independent simulation runs, and separate simulations are used for the
positive-going and negative-going scans. The dynamic CV has broader
peaks than the quasi-equilibrium CV obtained from the equilibrium
simulation.  The peaks are also displaced in the scan direction,
relative to the quasi-equilibrium peak positions.  Clearly, dynamic
simulations provide information about possible dynamic effects in CV
experiments with finite potential-scan rates.

Dynamic Monte Carlo simulations can also be used to study the current
transients that occur after the electrode potential is rapidly stepped
across a transition \cite{HOLZ94}.  In this case, the system is
equilibrated on one side of the phase transition (located by the peak
positions in the quasi-equilibrium CV), then the potential is changed
instantaneously across the transition to its final value.  The results
for potential steps simulations across the
$(\sqrt{3}\!\times\!\sqrt{3})_{2/3}^{1/3} \; \leftrightarrow
\;(1\!\times\!1)^0_0$ transition are shown in Fig.~\ref{fig:CuStep}.

Fig.~6a records current transients for potential steps that start
in the ordered phase and end between $24$ and $33$ mV past the
transition on the low-coverage side. The first current transient
corresponds to the relatively fast relaxation to a metastable state
with slightly less adsorbed Cu. This is followed be a second current
maximum as domains of empty surface nucleate and grow. The trend of
the second maximum to early times and larger currents as the step size
is increased agrees with recent experiments \cite{HOLZ94}.

The current transient for negative steps in the cell potential is
shown in Fig. 6b. The transient simply decreases with time,
corresponding to adsorption onto the empty surface that does not
require nucleation, {\it i.e.} the surface is unstable with respect to
adsorption after the potential step. This difference between the
observed transients after positive- and negative-going potential steps
across this particular transition is precisely the difference seen in
experiments of Cu with ${\rm SO}_4$ adsorbing onto Au(111)
\cite{HOLZ94}.

\subsection{Urea on Pt(100)}
\label{sec:urea}

The adsorption of urea onto Pt(100) surfaces requires an even more
complicated lattice-gas model than the two systems discussed above.
Experimental and simulated quasi-equilibrium CV currents for this
system are shown together in Fig.~\ref{fig:UreaCV}.  The urea coverage
$\Theta_{\rm U}$, measured {\em in situ\/} \cite{RIKV95,GAMB93B},
changes over a narrow potential range from near zero on the negative
side of the CV peak to approximately 1/4~ML on the positive side.
{\em Ex situ} LEED \cite{RIKV95} at potentials on the positive side of
the peak show an ordered $c(2\times4)$ structure, consistent with a
urea coverage of 1/4$\,$ML. On the negative side of the peak, only an
unreconstructed $(1\times1)$ surface with $\Theta_{\rm H}=1$ is seen
(H denotes hydrogen).

The model developed to account for these observations is based on the
assumption that urea, CO(NH$_2)_2$, coordinates the platinum through
its nitrogen atoms, with the C=O group pointing away from the surface.
Since the unstrained N-N distance in urea matches the lattice constant
of the square Pt(100) surface quite well, urea is assumed to 
occupy two nearest-neighbor adsorption sites on the square Pt(100)
lattice. Coulometry data indicate that the hydrogen
coverage changes from near unity on the negative-potential side of the 
CV peak to near zero over the same potential range where urea
becomes adsorbed. Therefore, hydrogen is assumed to adsorb in
the same on-top positions as the urea nitrogen atoms.
In the resulting model hydrogen is adsorbed at the nodes and urea on
the bonds of a square lattice. Simultaneous occupation by two or more
urea molecules of bonds that share a node is excluded, as is
occupation by hydrogen of a node adjacent to a bond occupied by urea.
An illustrative Monte Carlo snapshot of a typical $c(2\times4)$ 
configuration with thermal fluctuations is shown in Fig.~\ref{fig:Uconf}.

The configuration energies of this lattice-gas model are given by
Eq.~(\ref{eq:lg2}) with A=U (urea) and B=H (hydrogen). The effective
lattice-gas interactions were determined from ground-state
calculations followed by numerical Monte Carlo simulations of
quasi-equilibrium CV currents \cite{RIKV95,GAMB93B}. In order to
stabilize the observed $c(2\times4)$ phase, effective interactions
were included through $n=8$ \cite{RIKV95,GAMB93B}.

The numerical Monte Carlo simulations, which used systems with up to
$32\times32$ square-lattice unit cells, were performed with a
heat-bath cluster algorithm.  The clusters are cross-shaped regions,
consisting of five nearest-neighbor nodes plus their four connecting
bonds. After symmetry reductions, these clusters have 64 different
configurations. As a result, the code is relatively slow, measured in
CPU time per attempted update.  However, the algorithm effectively
removes energy barriers due to ``jamming'' configurations, especially
deep in the $c(2\times2)$ phase. In that phase region the
cross-cluster algorithm outperforms an algorithm based on a
minimal cluster consisting of one bond plus its two adjacent nodes by
at least a factor of several million in terms of CPU time to achieve
equilibration.  The good agreement of the quasi-equilibrium CV for the
lattice-gas model with the experimental results, which is shown in
Fig.~\ref{fig:UreaCV}, again demonstrates the utility of Monte Carlo
simulations in the study of electrochemical adsorption.

%
%

\section{Summary}
\label{sec:conclusion}

Our goal has been to convince practitioners of interfacial
electrochemistry that Monte Carlo simulation is a viable and useful
research tool, well within the computational means of most
laboratories. We have described how numerical Monte Carlo simulations
of lattice-gas models for specific electrochemical adsorption can
provide realistic estimates for a number of experimentally observable
quantities. These include strictly equilibrium ones such as coverages
and surface charges, quasi-equilibrium quantities like CV currents for
extremely slow potential-scan rates, and genuine nonequilibrium
quantities such as current transients and CV currents at high scan
rates. Information about model parameters, including effective lateral
interaction energies and diffusion barriers for adsorption/desorption
and lateral diffusion, can also be obtained. In this Chapter we have
presented detailed discussions of three specific systems. In doing
so, we have attempted to give enough detail about the simulations and
references to general texts on simulation that it should enable the
reader to write and use her or his own Monte Carlo simulations.

Most models with realistic geometric structure and interactions are
not exactly solvable, and the solution of even highly simplified models
must be accomplished by approximate means.  Monte Carlo simulation
represents a superior alternative to most mean-field based approaches
because it provides a much more accurate connection between the model
parameters and the values of the calculated observables. If comparison
of experimental and theoretical results is going to yield reliable
information about the values of microscopic model parameters, Monte
Carlo simulation should usually be the method of choice.

%
%

\section*{Acknowledgments}

We thank J.~X.\ Wang, T.~Wandlowski, and B.~M.\ Ocko for the
permission to use their data for bromine on Ag(100) in
Fig.~\ref{fig:BrIsotherm}, and J.~X.\ Wang for helpful comments.  We
acknowledge useful discussions on Monte Carlo simulation with
M. Kolesik.  P.~A.~R.\ is particularly happy to acknowledge his long
and fruitful collaboration with A.~Wieckowski.

This research was supported by Florida State University through the
Center for Materials Research and Technology and the Supercomputer
Computations Research Institute (DOE Contract No. DE-FC05-85ER-25000),
and by NSF Grant No. DMR-963483.


\begin{thebibliography}{10}

\bibitem{BLUM90}
L Blum. Adv Chem Phys 78:171--227, 1990.

\bibitem{EINS91}
See, {\it e.g.}, TL Einstein. Langmuir 7:2520--2527, 1991.

\bibitem{EINS95}
TL Einstein. In: WN Unertl, ed. Handbook of Surface Science, Vol.~2: Physical
Structure of Solid Surfaces. Amsterdam: Elsevier North-Holland, 1995.

\bibitem{BLUM96}
L Blum, DA Huckaby, M Legault. Electrochimica Acta 41:2207--2227, 1996.

\bibitem{RIKV91B}
PA Rikvold. Electrochimica Acta 36:1689--1694, 1991.

\bibitem{RIKV95}
PA Rikvold, M Gamboa-Aldeco, J Zhang, M Han, Q Wang, HL Richards, A
Wieckowski. Surf Sci 335:389--400, 1995.

\bibitem{RIKV96}
PA Rikvold, J Zhang, YE Sung, and A Wieckowski. Electrochimica Acta
41:2175--2184, 1996.

\bibitem{RIKV93D} 
A comparison of results of perturbative and nonperturbative
calculations for one and the same two-dimensional lattice-gas model is
given by 
PA Rikvold, MA Novotny. In: L Blum, FB Malik, ed. Condensed
Matter Theories, Vol.~8. New York: Plenum, 1993, pp. 627-635.

\bibitem{BIND92B}
K Binder, DW Heermann. Monte Carlo Simulation in Statistical
Physics. Berlin: Springer, 1992.

\bibitem{GOUL96}
H Gould, J Tobochnik. An Introduction to Computer Simulation Methods.
Reading, Massachusetts: Addison-Wesley, 1996.

\bibitem{BIND79} K Binder. In: K Binder, ed. Monte Carlo Methods in
Statistical Physics. 2nd ed. Berlin: Springer, 1986.

\bibitem{BIND92}
K Binder. Ann Rev Phys Chem 43:33--59, 1992.

\bibitem{OCKO97}
BM Ocko, JX Wang, T Wandlowski. Phys Rev Lett 79:1511--1514, 1997.

\bibitem{KOPE97}
MTM Koper. A lattice-gas model for halide adsorption on single-crystal
electrodes. Preprint, 1997.

\bibitem{WANG97}
JX Wang, T Wandlowski, BM Ocko. Electrochem Soc
Proc 97-17:293--301, 1997

\bibitem{HUCK90}
D Huckaby, L Blum. J Chem Phys 92:2646--2649, 1990.

\bibitem{BLUM91}
L Blum, DA Huckaby. J Chem Phys 94:6887--6894, 1991.

\bibitem{HUCK91}
DA Huckaby, L Blum. J Electroanal Chem 315:255--261, 1991.

\bibitem{ZHAN96}
J Zhang, YE Sung, PA Rikvold, A Wieckowski. J Chem Phys
104:5699--5712, 1996.

\bibitem{BLUM94A}
L Blum, DA Huckaby. J Electroanal Chem 375:69--77, 1994.

\bibitem{BLUM94B}
L Blum, M Legault, P Turq. J Electroanal Chem 379:35--41, 1994.

\bibitem{GAMB93B}
M Gamboa-Aldeco, P Mrozek, CK Rhee, A Wieckowski, Q Wang, PA Rikvold.
Surf Sci 297:L135--L140, 1993.

\bibitem{LEE79}
HH Lee, DP Landau. Phys Rev B 20:2893--2900, 1979.

\bibitem{HUCK84}
DA Huckaby, JM Kowalski. J Chem Phys 80:2163--2167, 1984.

\bibitem{RIKV88B}
PA Rikvold, JB Collins, GD Hansen, JD Gunton. Surf Sci
203:500--524, 1988.

\bibitem{SCHI81}
M Schick. Prog Surf Sci 11:245--292, 1981.

\bibitem{BARD80}
AJ Bard, LR Faulkner. Electrochemical Methods: Fundamentals and
Applications. New York: Wiley, 1980.

\bibitem{JENS98}
JA Jensen, KB Rider, M Salmeron, GA Somorjai. Phys Rev Lett
80:1228--1231, 1998.

\bibitem{RIKV91A}
PA Rikvold, MR Deakin. Surf Sci 249:180--193, 1991.

\bibitem{COLL89}
JB Collins, P Sacramento, PA Rikvold, JD Gunton. Surf Sci 221:277--298, 1989.

\bibitem{GLAU63}
RJ Glauber. J Math Phys 4:294--307, 1963.

\bibitem{CHAY97}
L Chayes, J Machta. Physica A 239:542--601, 1997.

\bibitem{varNote}
The factor $s/(s-1)$ in the variance in Eq.~(\protect\ref{eq:chi}) results
because both $M_1^{\rm X}$ and $M_2^{\rm X}$ are estimated from the same
sample. For a discussion of how to calculate empirical moments from numerical
data while controlling roundoff errors, see, {\em e.g.\/}, 
WP Press, SA Teukolsky, WT Vetterling, BP Flannery.
Numerical Recipes in Fortran 77, 2nd ed. Cambridge: Cambridge
University Press, 1996, p. 605.

\bibitem{PATHRIA}
RK Pathria. Statistical Mechanics. 2nd ed. Oxford: Butterworth-Heinemann, 1996.

\bibitem{GOLD92}
N Goldenfeld. Lectures on Phase Transitions and the Renormalization
Group. Reading, Massachusetts: Addison-Wesley, 1992, pp. 279--282.

\bibitem{URL}
Monte Carlo movies and other supplementary materials can be viewed at {\tt
http://www.scri.fsu.edu/\~{}rikvold}~.

\bibitem{VATT96}
I Vattulainen, J Merikoski, T Ala-Nissil{\"a}, SC Ying. Surf Sci
366:L697--L702, 1996.

\bibitem{VATT92}
T Ala-Nissil{\"a}, J Kjoll, SC Ying. Phys Rev B 46:846--854, 1992.

\bibitem{BORT75}
AB Bortz, MH Kalos, JL Lebowitz. J Comput Phys 17:10--18, 1975.

\bibitem{NOVO95}
MA Novotny. Computers in Physics 9:46--52, 1995.

\bibitem{AHO}
AV Aho, JE Hopcroft, JD Ullman. Data Structures and
Algorithms. Reading, Massachusetts: Addison-Wesley, 1983.

\bibitem{VALE78}
G Valette, A Hamelin, R Parsons. Z Phys Chem 113:71--89, 1978.

\bibitem{BAXT80}
RJ Baxter, IG Enting, SK Tsang. J Stat Phys 22:465--489, 1980.

\bibitem{GIAC95}
G Giacomin, JL Lebowitz, C Maes. J Stat Phys 80:1379--1403, 1995.

\bibitem{RACZ80}
Z Racz. Phys Rev B 21:4012--4016, 1980.

\bibitem{KINZ81}
W Kinzel, M Schick. Phys Rev B 24:324--328, 1981.

\bibitem{BIND80}
K Binder, DP Landau. Phys Rev B 21:1941--1962, 1980.

\bibitem{SCHU76}
JW Schultze, D Dickertmann. Surf Sci 54:489--504, 1976.

\bibitem{DICK76}
D Dickertmann, FD Koppitz, JW Schultze. Electrochim Acta 21:967--971, 1976.

\bibitem{KOLB87}
DM Kolb. Z Phys Chem Neue Folge 154:179--199, 1987.

\bibitem{ZEI87}
M Zei, G Qiao, G Lempfuhl, DM Kolb. Ber Buns Ges Phys Chem 91:3494, 1987.

\bibitem{TONE95}
MF Toney, JN Howard, H Richer, GL Borges, JG Gordon, OR Melroy, D Yee,
LB Sorensen. Phys Rev Lett 75:4472--4475, 1995.

\bibitem{RIKV97A}
PA Rikvold, G Brown, MA Novotny, A Wieckowski. Coll Surf A 134:3--14, 1998.

\bibitem{BAXTER}
RJ Baxter. Exactly Solved Models in Statistical Mechanics. 
London: Academic Press, 1982.

\bibitem{MAGN90}
OM Magnussen, J Hotlos, RJ Nichols, DM Kolb, RJ Behm. 
Phys Rev Lett 64:2929--2932, 1990.

\bibitem{EDEN94}
GJ Edens, X Gao, MJ Weaver. J Electroanal Chem 375:357--366, 1994.

\bibitem{OMAR93}
IH Omar, HJ Pauling, K J{\"u}ttner. J Electrochem Soc 140:2187--2192, 1993.

\bibitem{ZSHI94C}
Z Shi, J Lipkowski. J Electroanal Chem 365:303--309, 1994.

\bibitem{ZSHI95B}
Z Shi, J Lipkowski, S Mirwald, B Pettinger. J Electroanal Chem 396:115--124, 1995.

\bibitem{HOLZ94}
MH H{\"o}lzle, U Retter, DM Kolb. J Electroanal Chem 371:101--109, 1994.

\end{thebibliography}

%
%

~
\begin{figure}[ht]
\caption[]{ Ground-state diagram for the lattice-gas model of copper
UPD on Au(111), shown in the ($\bar{\mu}_{\rm S} , \bar{\mu}_{\rm C}$)
plane. The subscripts S and C denote sulfate and Cu, respectively.
The effective interactions are given in Ref.~\protect\cite{ZHAN96}.
The solid lines represent zero-temperature phase boundaries, the
dotted line is a zero-temperature phase boundary that is not observed
in room-temperature simulations. The voltammetric scan path for the
simulations discussed here is represented as the dashed line.  Its end
points correspond to $E$=110~mV (upper left) and 300~mV (lower right) vs
Ag/AgCl, respectively. The phases are denoted as $(X \! \times \!
Y)_{\Theta_{\rm C}}^{\Theta_{\rm S}}$. The ground-state
configurations corresponding to the main phases are also shown. The
adsorption sites are shown as {\large $\circ$}, and Cu and sulfate are
denoted by {\large $\bullet$} and $\triangle$, respectively.  (Adapted from
Ref.~\protect\cite{ZHAN96}.)  }
\label{fig:gstate}
\end{figure}
\vfill 

~
\begin{figure}[tbp]
\caption{Schematic free-energy surfaces for thermally activated
transitions in the dynamic Monte Carlo model. The text describes how
these free-energy surfaces, based on the Butler-Volmer model, are
used to calculate energy barriers for Monte Carlo processes. In our
simulations processes corresponding to adsorption, desorption, and
surface diffusion are included.}
\label{fig:Barrier}
\end{figure}
\vfill

~
\begin{figure}[tbp]
\caption[]{Snapshots of configurations from simulation of bromine
adsorption on Ag(100) for (a) the low-coverage disordered phase and
(b) the ordered $c(2 \times 2)$ phase. The light gray circles
represent surface Ag atoms, and the black circles the adsorbed Br
atoms. The Br adsorb to the four-fold hollow sites, which form a
square lattice.}
\label{fig:BrSnap}
\end{figure} 
\vfill

~
\begin{figure}[tbp]
\caption[]{Experimental and theoretical isotherms for bromine
adsorption on Ag(100). The circles are chronocoulometry results from
Ref.~\protect{\cite{OCKO97}}. The solid curve is the numerical
Monte Carlo result, while the dashed curve is the best-fit
single-sublattice Frumkin isotherm. The Monte Carlo isotherm displays
the singularity associated with the phase transition, while the
Frumkin isotherm does not.}
\label{fig:BrIsotherm}
\end{figure}
\vfill

~
\begin{figure}[ht]
\caption[]{ Simulated CV profiles for copper UPD on Au(111),
corresponding to the scan path in Figure~\protect\ref{fig:gstate}.
Solid curves are dynamic simulations with scan rate $10\,{\rm mV/s}$,
while dotted curves are quasi-equilibrium estimates. The model
parameters are the same as in Ref.~\protect\cite{ZHAN96}.}
\label{fig:UPDcv}
\end{figure}
\vfill

~
\begin{figure}[tbp]
\caption[]{Simulated current transients after potential steps across
the $(1\!\times\!1)_0^0 \leftrightarrow
(\sqrt{3}\!\times\!\sqrt{3})^{1/3}_{2/3}$ transition. (a) Positive
steps. (b) Negative step. The observed asymmetry is the same in the
simulations as in experiments.  (Reproduced with permission from
Ref.~\protect\cite{RIKV95}.  Copyright 1998 Elsevier Science.)}
\label{fig:CuStep}
\end{figure}
\vfill

~
\begin{figure}[ht]
\caption[]{ Room-temperature CV profiles for urea on Pt(100) in
0.1$\,$M HClO$_4$.  Experimental (dashed curves) and simulated
($\Diamond$ and solid curve) quasi-equilibrium normalized CV currents,
$i$/(d$E$/d$t$), in elementary charges per mV per Pt(100) unit cell,
at 1.0$\,$mM bulk urea. The dashed curves are representative
negative-going voltammograms; two are shown to demonstrate variations
between individual measurements. The effective interactions and
electrovalences are given in the caption of Fig.~4 of
Ref.~\protect\cite{GAMB93B}.  The simulation results shown here are
for a system of $32\times32$ Pt(100) unit cells.  (Reproduced with
permission from Ref.~\protect\cite{RIKV95}.  Copyright 1995 Elsevier
Science.)  }
\label{fig:UreaCV}
\end{figure}
\vfill

~
\begin{figure}[ht]
\caption[]{ Left: The microscopic model for urea on Pt(100), showing
urea molecules with their NH$_2$ groups (dark gray) on the square
Pt(100) lattice sites and their C=O groups (C black and O lighter
gray) pointing away from the surface. The hydrogen atoms are shown as
single light gray spheres.  Right: A typical equilibrium configuration
generated by Monte Carlo in the ordered $c(2\times4)$ phase at
$-$60$\,$mV, using the model parameters given in
Ref.~\protect\cite{RIKV95}.  Urea molecules are shown as filled
rectangles on the lattice bonds and hydrogen as $\bullet$ at the
nodes.  (Reproduced with permission from Ref.~\protect\cite{RIKV95}.
Copyright 1995 Elsevier Science.)  }
\label{fig:Uconf}
\end{figure}
\vfill

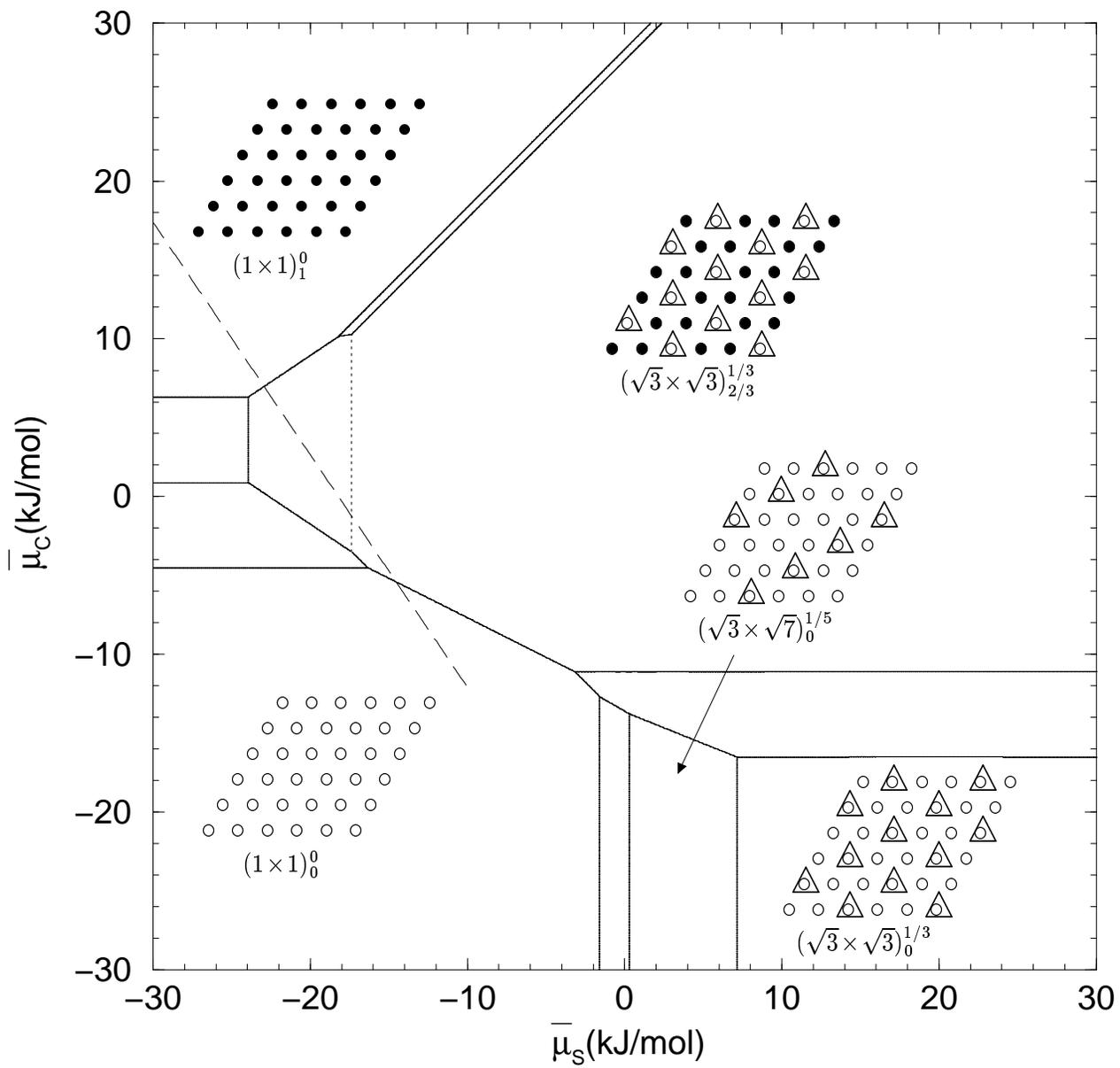
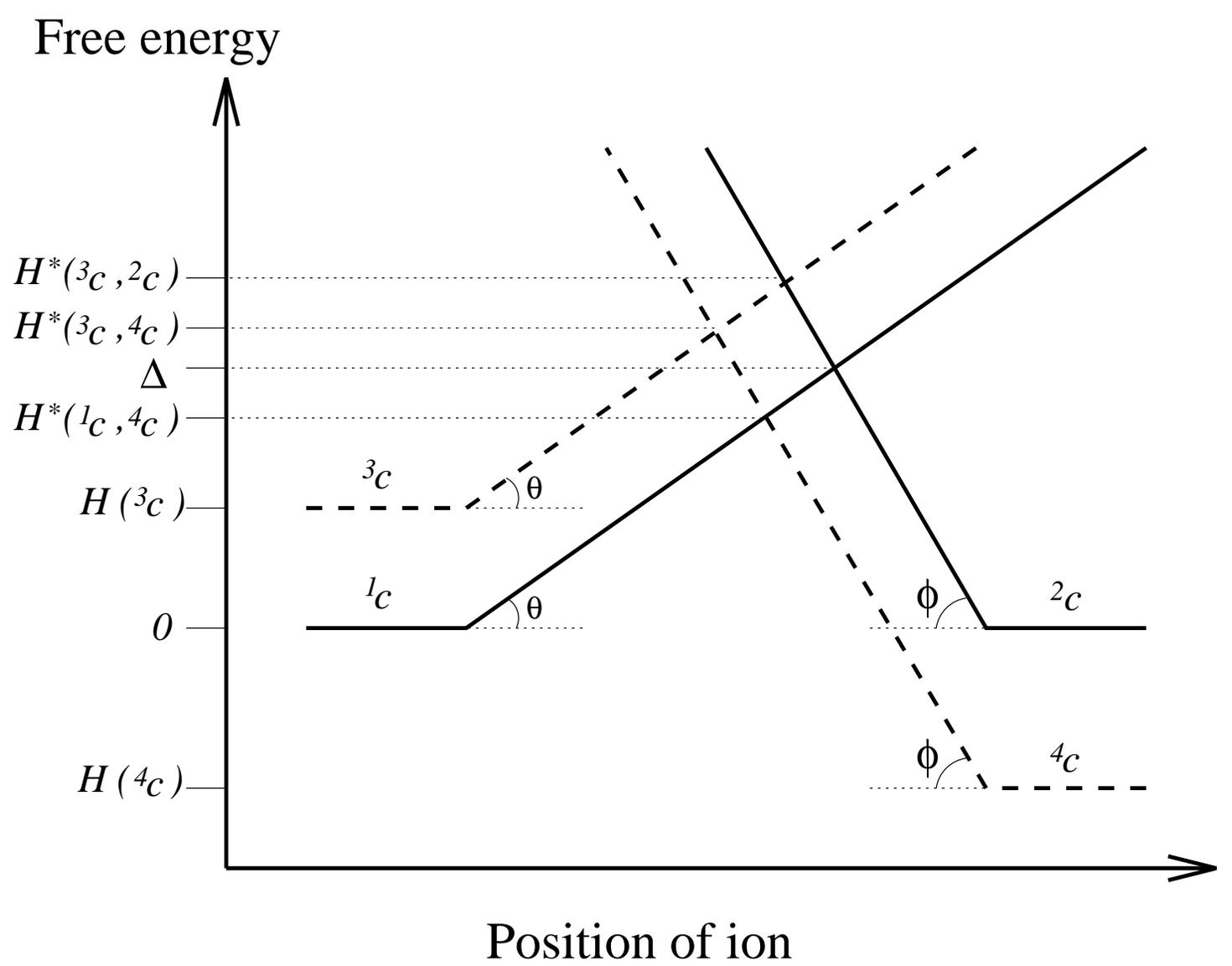
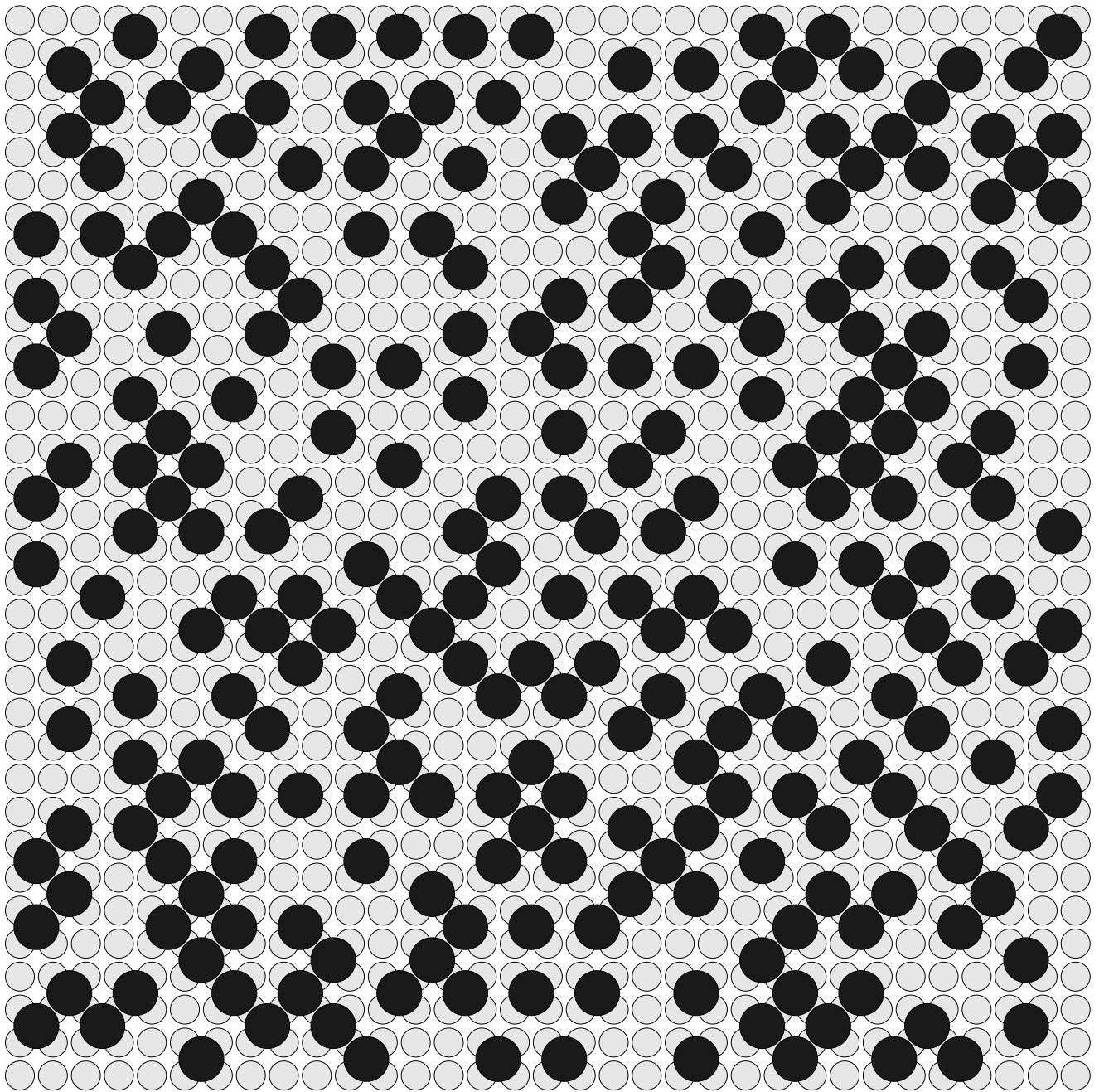
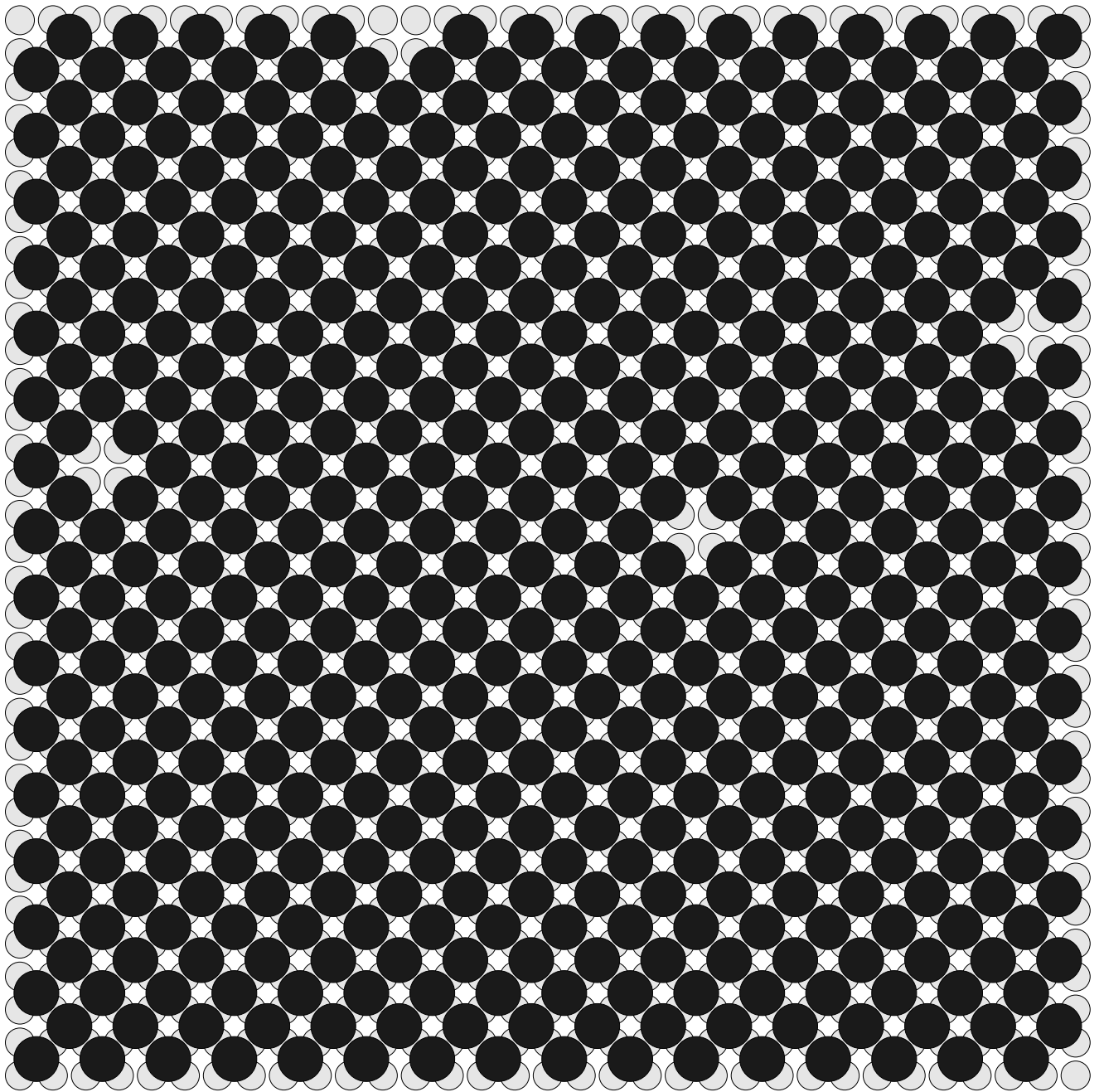
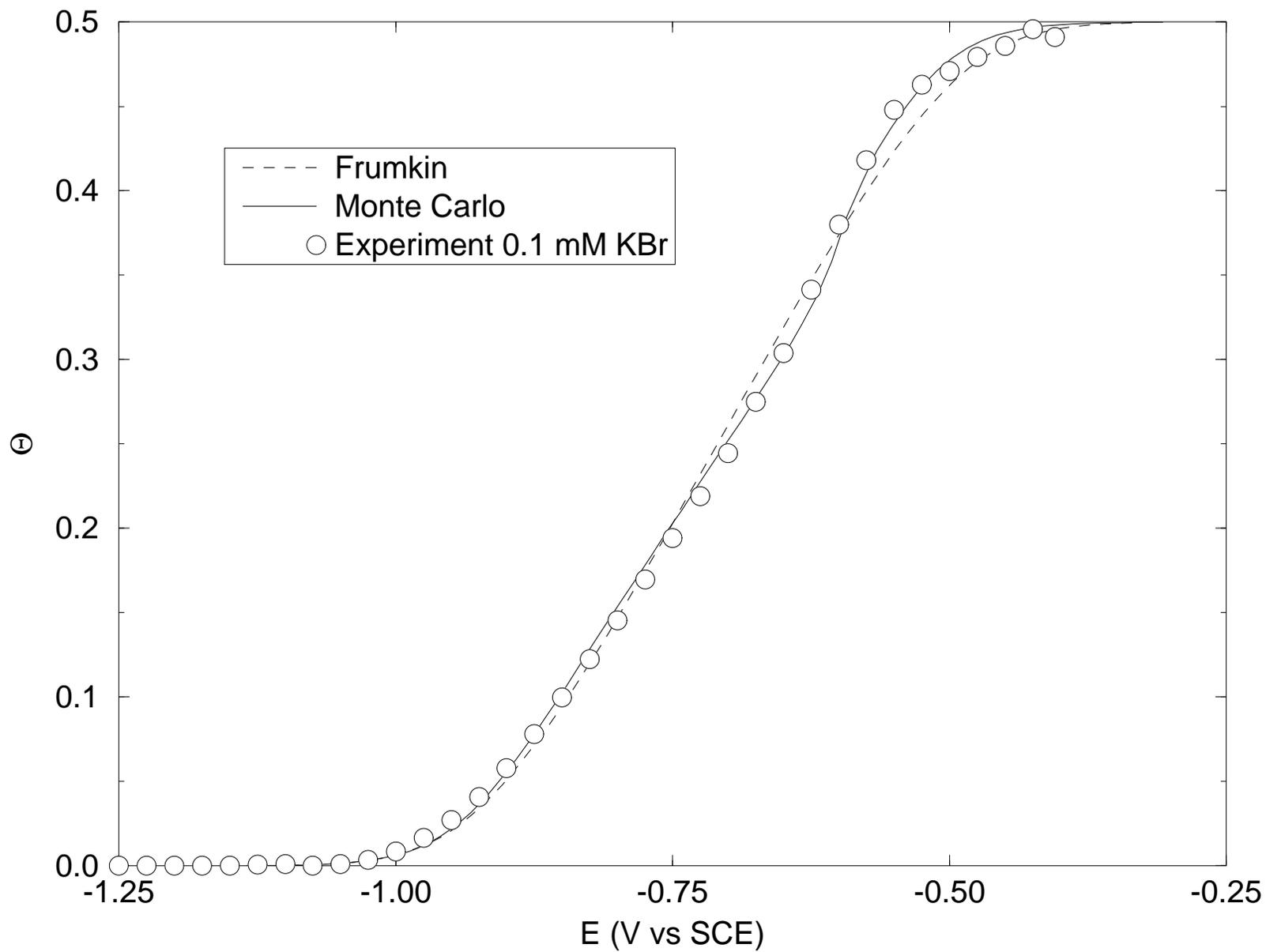
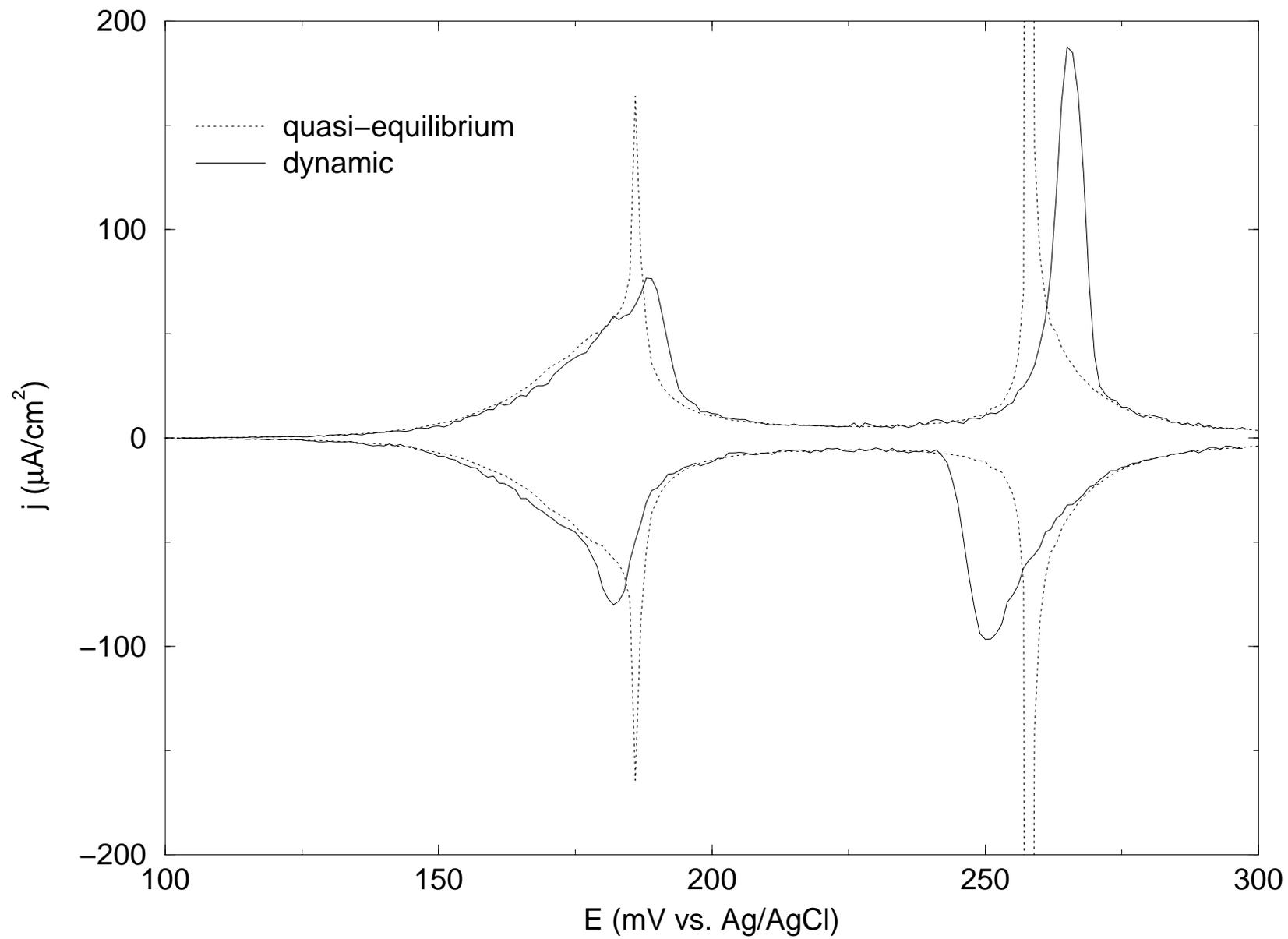
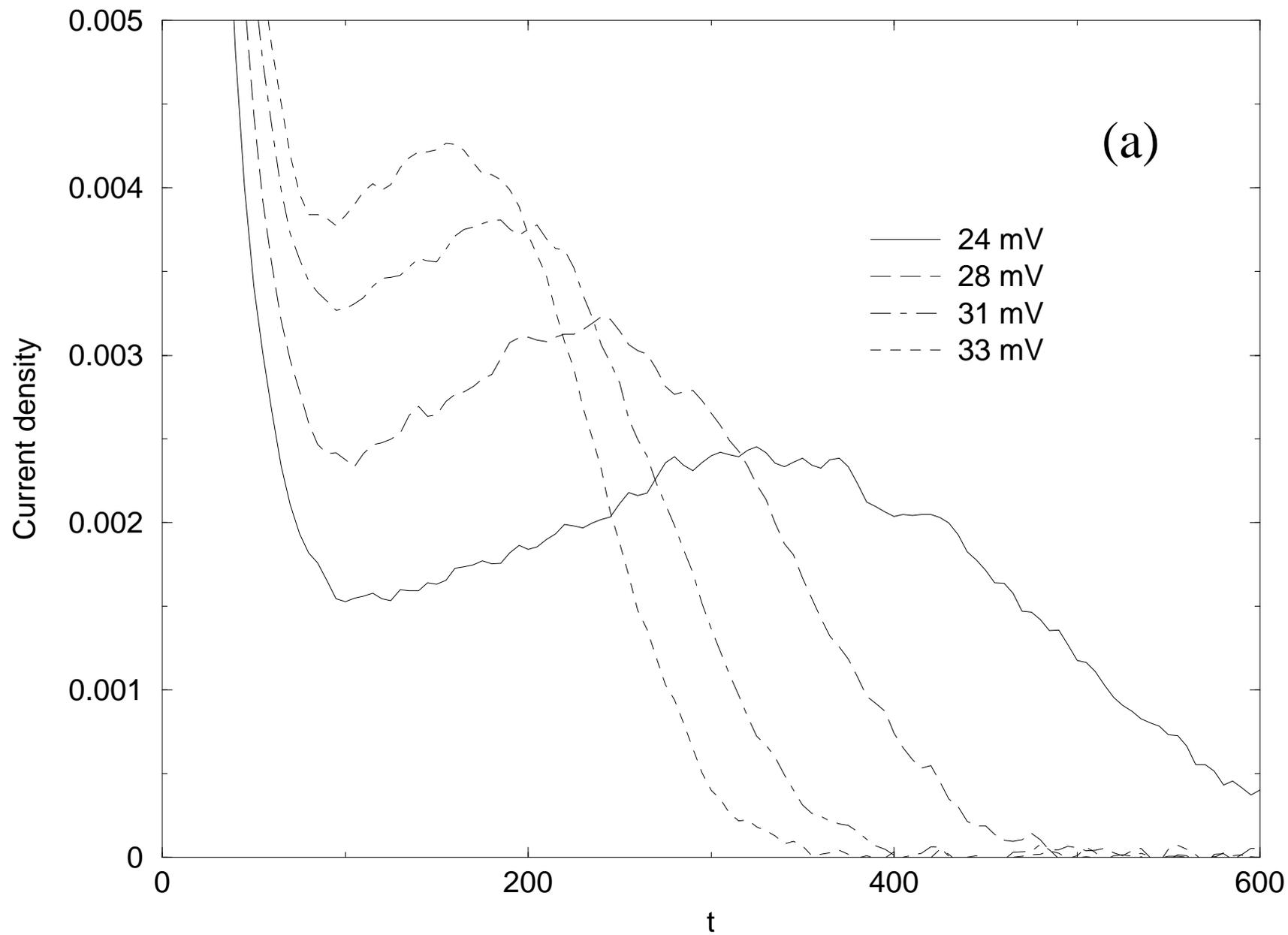
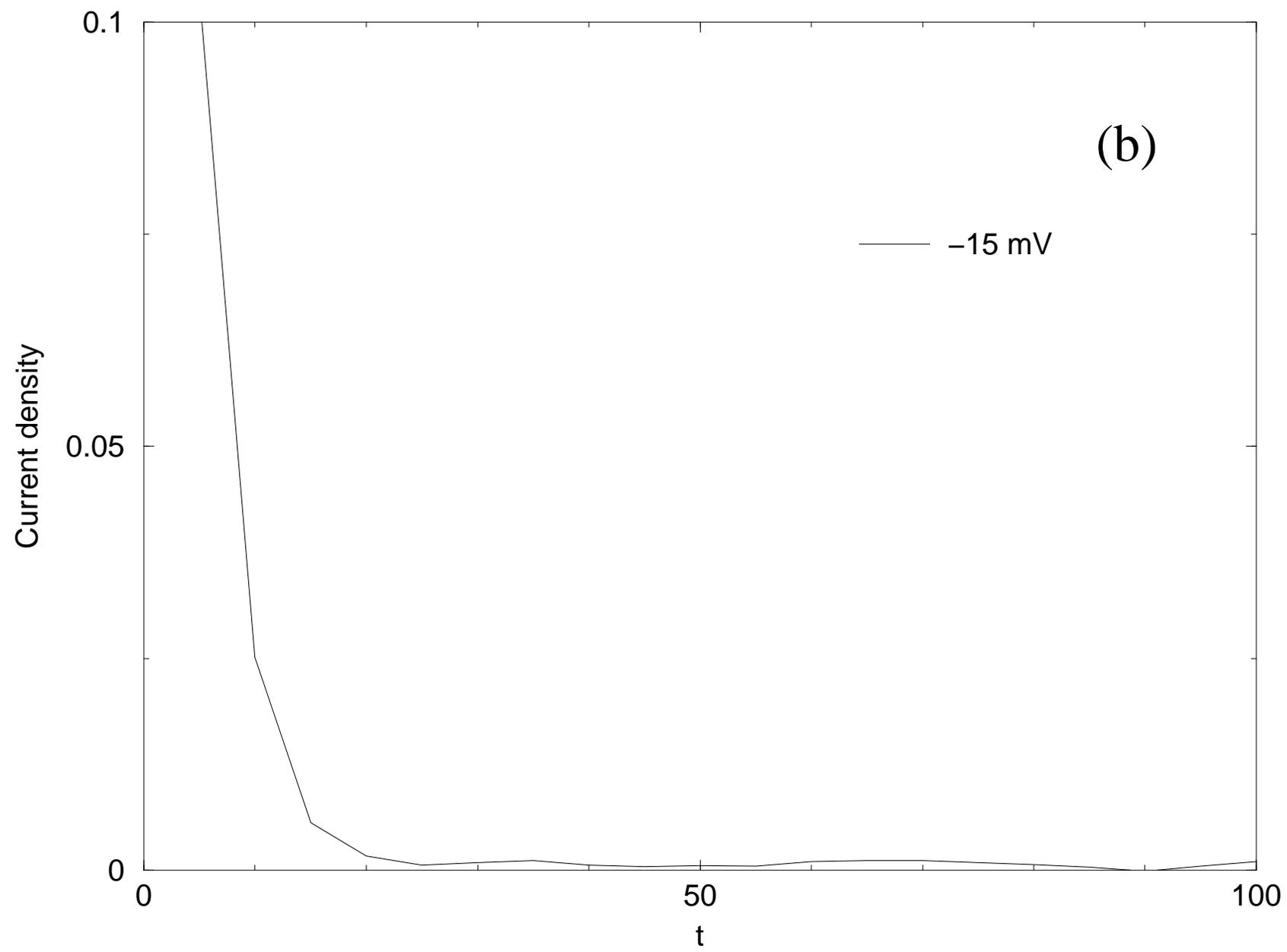
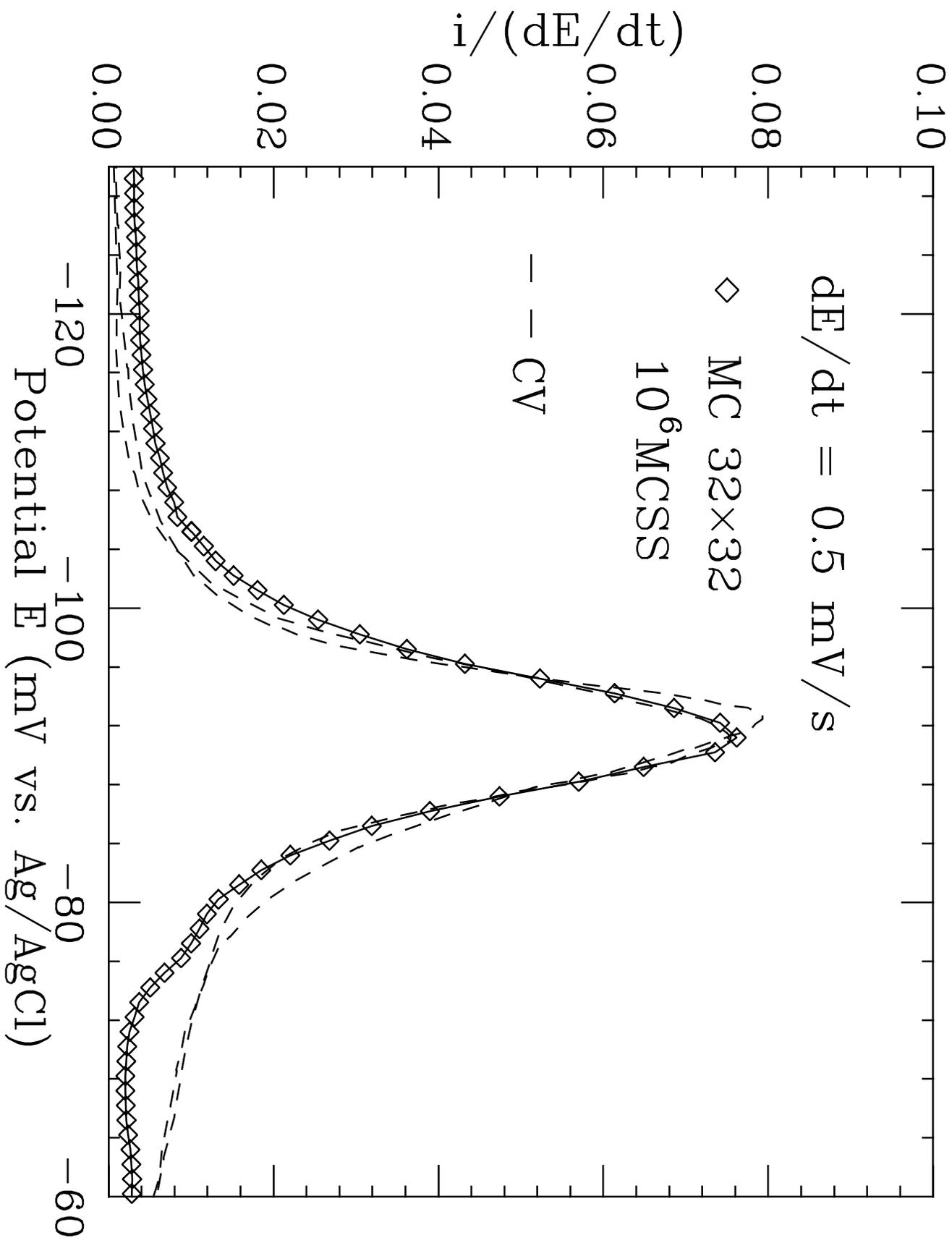
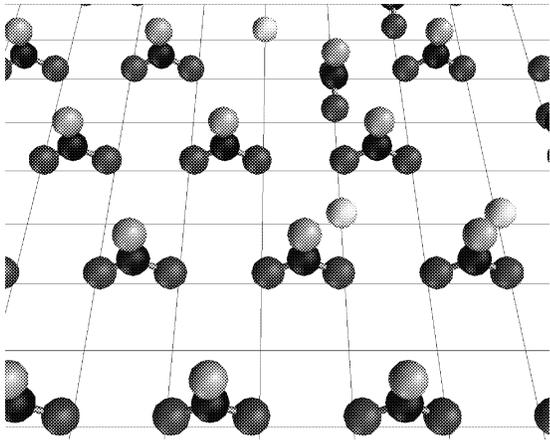
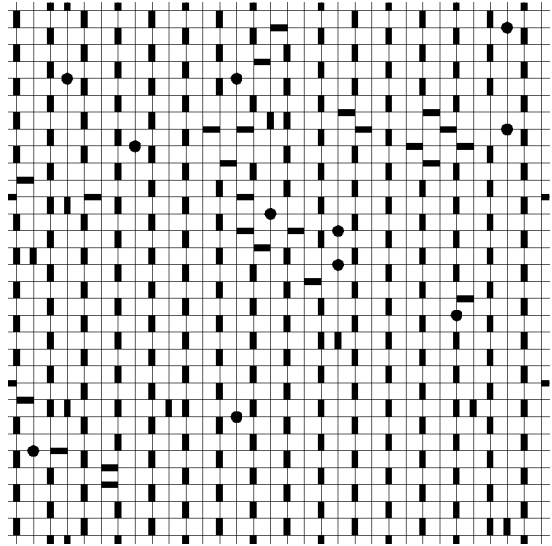
\end{document}